\documentclass[11pt]{article}
\usepackage{epsfig,amsmath, amsfonts}

\newcommand{\blank}[1]{}
\def\clim{{\ensuremath{c \to 1}}}
\def\plim{{\ensuremath{p \to \infty}}}
\def\xlim{{\ensuremath{x \to 0}}}
\def\Moo{{\ensuremath{M_\infty}}}
\def\D{{\mathrm d}}
\def\R{{r}}
\def\L{{\ell}}

\newcommand{\B}[3]{\,\vphantom{B}^{#1}{\!B_{#2}}^{#3}\,}
\newcommand{\bnpt}[3]{\ensuremath{\langle{#1}\rangle^{\!#2\!}_{\text{\tiny{#3}}}}}
\newcommand{\C}[3]{\ensuremath{\,{C_{{#1}{#2}}}^{#3}\,}}

\newcommand{\eps}{\ensuremath{\varepsilon}}

\newcommand{\G}[1]{\Gamma\!\big(#1\big)\,}
\newcommand{\Hc}{\mathcal{H}}
\newcommand{\hf}{{{\scriptstyle\frac{1}{2}}}}
\newcommand{\Id}{{1\hspace{-4.5pt}1}}
\newcommand{\is}[1]{|#1\rangle}
\newcommand{\ishin}[1]{\ensuremath{|#1\rangle\!\rangle}}
\newcommand{\ishout}[1]{\ensuremath{\langle\!\langle{#1}|}}
\renewcommand{\l}{\ell}
\newcommand{\Lb}{\ensuremath{\bar L}}
\newcommand{\npt}[1]{\ensuremath{\langle{#1}\rangle}}
\newcommand{\os}[1]{\langle{#1}|}
\newcommand{\qt}{{\tilde q}}
\newcommand{\Rbb}{\mathbb{R}}
\renewcommand{\S}[2]{\ensuremath{{S_{#1}}^{#2}}}
\newcommand{\Sbb}{{\ensuremath{\mathbb{S}}}}
\newcommand{\tf}{{{\scriptstyle\frac{3}{2}}}}

\newcommand{\ze}{\zeta}
\newcommand{\zeb}{{\bar \zeta}}
\newcommand{\Zbb}{\mathbb{Z}}

\newcommand{\blockA}[3]{\ensuremath{
     \setlength{\unitlength}{0.75mm}
     \begin{picture}(10,15)(0,0)
          \put( 0, 4){\line(1,0){10}}
          \put(10, 4){\line(0,1){10}}
          \put( 2, 5){\makebox(0,0)[lb]{\small $#1$}}
          \put(11,13){\makebox(0,0)[lt]{\small $#2$}}
          \put(10, 3){\makebox(0,0)[ct]{\small $#3$}}
     \end{picture}}}

\newcommand{\blockB}[1]{\ensuremath{
     \setlength{\unitlength}{0.75mm}
     \begin{picture}(10,15)(0,0)
          \put( 0, 4){\line(1,0){10}}
          \put( 8, 5){\makebox(0,0)[rb]{\small $#1$}}
     \end{picture}}}

\newcommand{\blockC}[3]{\ensuremath{
     \setlength{\unitlength}{0.75mm}
     \begin{picture}(10,15)(0,0)
          \put( 0, 4){\line(1,0){10}}
          \put(10, 4){\line(0,1){10}}
          \put( 5, 5){\makebox(0,0)[cb]{\small $#1$}}
          \put(11,13){\makebox(0,0)[lt]{\small $#2$}}
          \put(10, 3){\makebox(0,0)[ct]{\small $#3$}}
     \end{picture}}}

\newcommand{\cbB}[7]{\blockA{#1}{#2}{#6}\blockC{#5}{#3}{#7}\blockB{#4}}
\newcommand{\bL}[1]{\raisebox{-2mm}{#1}}

\newcommand{\xfiga}{\begin{picture}(0,0)%
\includegraphics{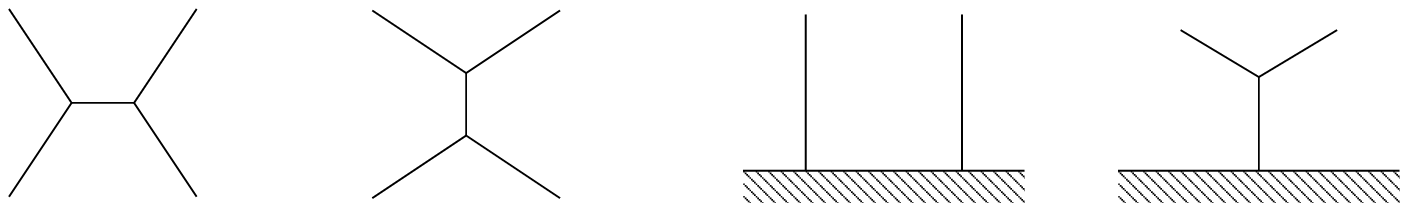}%
\end{picture}%
\setlength{\unitlength}{3947sp}%
\begin{picture}(7212,1081)(1,-362)
\put(3751,539){\makebox(0,0)[lb]{\smash{b)}}}
\put(  1,539){\makebox(0,0)[lb]{\smash{a)}}}
\put(1471,539){\makebox(0,0)[lb]{\smash{$y$}}}
\put(1471,-361){\makebox(0,0)[lb]{\smash{$z$}}}
\put(496,539){\makebox(0,0)[rb]{\smash{$x$}}}
\put(496,-361){\makebox(0,0)[rb]{\smash{$t$}}}
\put(3196,539){\makebox(0,0)[lb]{\smash{$y$}}}
\put(3196,-361){\makebox(0,0)[lb]{\smash{$z$}}}
\put(2221,539){\makebox(0,0)[rb]{\smash{$x$}}}
\put(2221,-361){\makebox(0,0)[rb]{\smash{$t$}}}
\put(2751, 89){\makebox(0,0)[lb]{\smash{$q$}}}
\put(903,199){\makebox(0,0)[lb]{\smash{$p$}}}
\put(4726,-136){\makebox(0,0)[b]{\smash{$1$}}}
\put(4261,464){\makebox(0,0)[b]{\smash{$x$}}}
\put(5236,464){\makebox(0,0)[b]{\smash{$y$}}}
\put(6046,464){\makebox(0,0)[b]{\smash{$x$}}}
\put(6642,  7){\makebox(0,0)[b]{\smash{$p$}}}
\put(7024,464){\makebox(0,0)[b]{\smash{$y$}}}
\end{picture}}

\begin{document}

\begin{titlepage}
\vskip 0.5cm
\begin{flushright}
KCL-MTH-01-23\\
PAR-LPTHE-01-32\\
{\tt hep-th/0107118}\\
\end{flushright}
\vskip 4.cm
\begin{center}
{\Large\bf A non-rational CFT with c=1\\[10pt]
 as a limit of minimal models } \\[5pt]
\end{center}
\vskip 1.3cm
\centerline{I.~Runkel\footnote{e-mail: {\tt ingo@lpthe.jussieu.fr}}
and G.M.T.~Watts\footnote{e-mail: {\tt gmtw@mth.kcl.ac.uk}}
}
\vskip 0.6cm
\centerline{${}^1$\sl LPTHE, Universit\'e Paris VI,}
\centerline{\sl 4 place Jussieu, F\,--\,75\,252\, Paris\, Cedex 05, France}
\vskip 0.4cm
\centerline{${}^2$\sl Mathematics Department, }
\centerline{\sl King's College London, Strand, London WC2R 2LS, U.K.}
\vskip 0.9cm
\begin{abstract}
\vskip0.15cm
\noindent
  We investigate the limit of minimal model conformal field theories
where the central charge approaches one. We conjecture that this limit
is described by a non-rational CFT of central
charge one. The limiting theory is different from the free boson but
bears some resemblance to Liouville theory. 
Explicit expressions for the three point functions of 
bulk fields are presented, as well as a set of conformal boundary
states. We provide analytic and numerical arguments in support of the
claim that this data forms a consistent CFT.

\end{abstract}
\end{titlepage}
\setcounter{footnote}{0}
\def\thefootnote{\fnsymbol{footnote}}

\section{Introduction}

The unitary minimal models were discovered in \cite{BPZ84}. They are
two dimensional conformal field theories with the special property
that they only contain a finite number of primary fields with respect
to the Virasoro algebra.  The possible partition functions for these
models were classified in \cite{CIZ87},  and there is one unitary
minimal model $M_p$ for each integer $p\ge 2$ with a diagonal modular
invariant.  (Throughout this paper we will only consider only theories
with diagonal modular invariant.)  $M_p$ has central charge
\begin{align}
  c = 1 - \frac{6}{p(p{+}1)} \;.
\end{align}
The first non-trivial minimal model, $M_3$, with central charge $c=1/2$, 
corresponds to the critical Ising model. At the other end, for
\plim, we see that the central charge approaches the
limiting value $c=1$. 

Each of the minimal models has a set of conformal boundary conditions
and in \cite{GRW01} it was shown how a particular limit of these
boundary conditions is useful in clarifying the space of boundary
renormalisation group flows in the minimal models.
In that paper, 
only the boundary theories were considered, and 
it was found that at the limit point $c=1$ there is an infinite
discrete set of fundamental conformal boundary conditions, labelled by
positive integers $a$. All other conformal boundary conditions
obtained in the \clim\  limit can be expressed as superpositions of
these fundamental ones. 

It appears that none of the known $c=1$ theories has such a set of
boundary conditions, and so it is natural to ask if it is possible
to define a `limiting minimal model'
\begin{align}
  \Moo = \lim_{\plim} M_p \;,
  \label{intro:Minf}
\end{align}
such that the boundary conditions of \Moo\
are those discussed in \cite{GRW01}.
It seems probable that one may take the limit \plim\ of the minimal
models in more than one way \cite{RWp.c.}, but we shall not be
concerned with such issues here and concentrate on finding a bulk
field theory associated to the boundary conditions discussed
in \cite{GRW01}.

Each minimal model $M_p$ has a finite number of Virasoro primary
bulk fields, and requiring that the boundary conditions remain a
discrete set each with a finite number of boundary fields 
implies that the bulk spectrum becomes continuous in the limit \plim.
\Moo\ thus contains a continuum of primary bulk fields of
conformal weight $h$. For later convenience we shall parametrise these
fields in terms of $x\,{=}\,2 \sqrt h$ (so that $h_x\,{=}\,x^2/4$) and
denote them  by $\phi_x(\ze,\zeb)$
(n.b. $h=\bar h$ for all primary fields in \Moo).

Curiously, the resulting bulk theory for \Moo\ is 
not that of a free boson. In the case of the free boson,
conservation of the $U(1)$ charge strongly constrains the operator
product expansion of two vertex operators. 
If the vertex operator $V_x$ has charge $x$, then only the one vertex
operator with correct charge can appear on the right-hand side of the
OPE:  
\begin{align}
      V_x(\ze,\zeb) \, V_y(0,0) 
\;=\;
      |\zeta|^{xy}\,
      \big( V_{x+y}(0,0)
      \;+\; \hbox{descendants} \big) 
\;.
\end{align}
For \Moo\ however there is no corresponding conserved charge and
one finds a continuum of primary fields when taking the OPE:
\begin{align}
  \phi_x(\ze,\zeb) \,
  \phi_y(0,0) 
\;=\;
   \int 
  |\zeta|^{2h_z - 2 h_x - 2 h_y}\,
  c(x,y,z)\,
  \big( \phi_z(0,0)\;+\; \hbox{descendants} \big) \;\D z \;.
  \label{intro:Minf-ope}
\end{align}
In this respect \Moo\ is rather similar to Liouville theory
\cite{DOt94,ZZa95,Tes01}, 
and indeed the relation of this theory to
Liouville theory is a major unanswered question.

A conformal field theory is rational (with respect to a chiral
algebra) if it contains a finite number of primary fields with respect
to that algebra, and quasi-rational (a weaker condition) if the
operator product expansion of any two primary fields contains
contributions from only a finite number of primary fields.
A free boson is not rational with respect to either the Virasoro
algebra or the larger $U(1)$-current algebra, but is quasi-rational. 
\Moo\ on the other hand is, just like Liouville theory, neither
rational nor quasi-rational.

The main result of this paper is the explicit formula for the
\Moo\ structure constants $c(x,y,z)$ appearing in the OPE
\eqref{intro:Minf-ope}. This is given in section \ref{sec:2.1}.
We are able to test these structure constants by a check of crossing
symmetry of the bulk four-point functions, both analytically (in the
special case when the $c=1$ chiral blocks are known) and numerically
(in the general case).
We also propose a set of boundary states which correspond to the
discrete set of
$c=1$ boundary conditions found in \cite{GRW01}. 

The paper is divided in three parts. First the $c=1$ theory \Moo\ is
presented, in terms of its two-- and three point functions on the
complex plane, and in terms of boundary states and one point functions
on the unit disc. Next, the correlators of $M_\infty$
are worked out in terms of conformal blocks and 
numerical and analytic
tests of crossing symmetry are described. The final section deals
with the derivation of the \Moo\ quantities presented before.


\section{Properties of $M_\infty$}
\label{sec:c=1}

In this section we describe in some detail the conjectured $c=1$
theory, which was denoted by \Moo\ in the introduction. 
All expressions given here have been obtained as the limits of
the corresponding quantities in minimal models. For clarity of
presentation, the somewhat technical calculations of these limits have
been shifted to section \ref{sec:lim}.


\subsection{\Moo\ on the complex plane}
\label{sec:2.1}

For a unitary conformal field theory on the full complex plane all
correlators are specified once the two-- and three point functions of
the Virasoro
primary bulk fields are known (how this works in the example of
the four point function is shown in some detail in section \ref{sec:cross}).
Conformal invariance implies that all
fields except for the identity will have vanishing one point functions on
the complex plane. The functional dependence of two-- and three point
functions is also fixed by conformal invariance, at least up to a
constant. In the two point functions we can choose this constant at
our convenience by redefining the primary fields. 
The only non trivial input specifying the theory on the complex plane
is thus the primary fields 
which are present and the values
of the numerical constants appearing in the three point functions. 

We propose that the Hilbert space of \Moo\ contains 
highest weight
states $\is{x}$ of conformal weights $h_x=\bar h_x=x^2/4$
for all positive $x$ with the exception of the positive integers
i.e. the set 
\begin{align}
  \Sbb = \Rbb_{>0} - \Zbb_{>0} 
\;.
\label{c=1:spec}
\end{align}
Corresponding to these states there are bulk primary fields
\begin{align}
  \phi_x(\ze,\bar\ze)
\;,
\end{align}
of the same conformal weights.
Note that the fields taken out of \Sbb\ correspond exactly to those
values of $h$ for which the $c=1$ Virasoro highest weight
representations have null vectors (see\ section \ref{sec:repn}).

Among the weights absent from \Sbb\ is $h=0$, corresponding to the
identity operator and its
descendants (including $T(z)$ and $\bar T(\bar z)$).
Naturally we do not want to exclude these from the spectrum of fields,
but it does appear that there is no actual vacuum state $\is{0}$ in the
theory%
\footnote{%
In equation \eqref{eq:vacs} we define states $\is{0}$ and $\os{0}$ as
limits of states in $\Sbb$, but they are not normalisable}. 
In this case, the reconstruction of correlators
involving the identity field and its descendants is possible using the 
conformal Ward identities (expressing the commutation relations of the
Virasoro algebra with primary fields).
However, we are also able to show that the operator 
$\lim_{x \to 0}\tfrac{d}{dx} \phi_x(\ze,\zeb)$ behaves exactly as the
identity field, and so in this way the identity can be constructed as
a limit of the fields in \Sbb.
Similarly, while it may be possible to construct fields
for the other integer values of $x$ in terms of the $\phi_x$, there
are no contributions from states with these weights to any correlation
function of fields in $\Sbb$, 
and so we have not detected their presence in the limit
theory we define.  We discuss this further in sections
\ref{sec:unprop} and \ref{sec:novac}. 

\begin{figure}[tb]
\vspace{-1cm}
\begin{tabular}{ll}
\epsfig{file=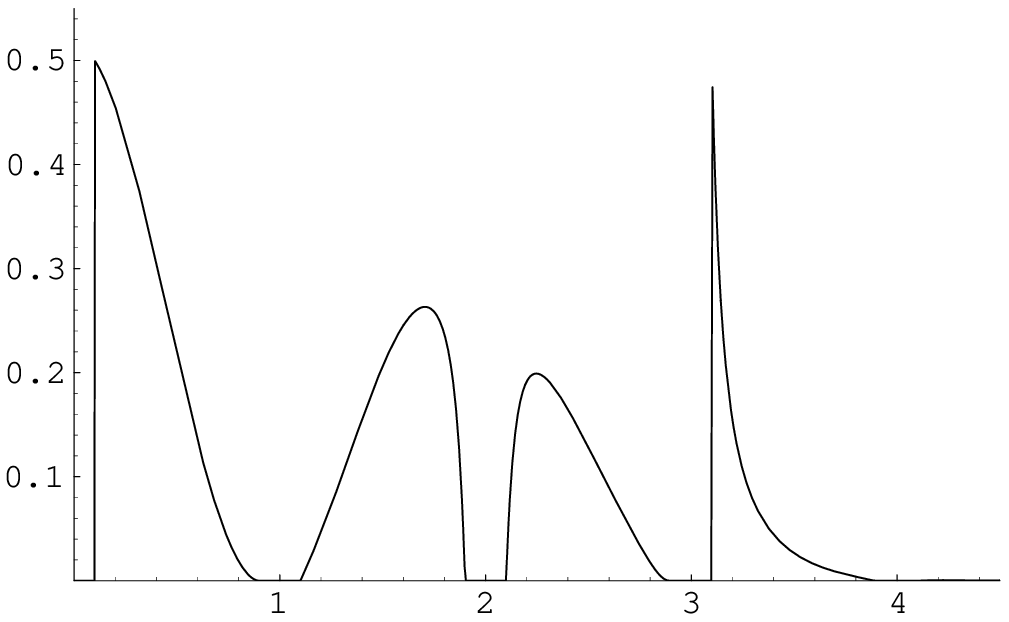, width=0.45\linewidth} &
\epsfig{file=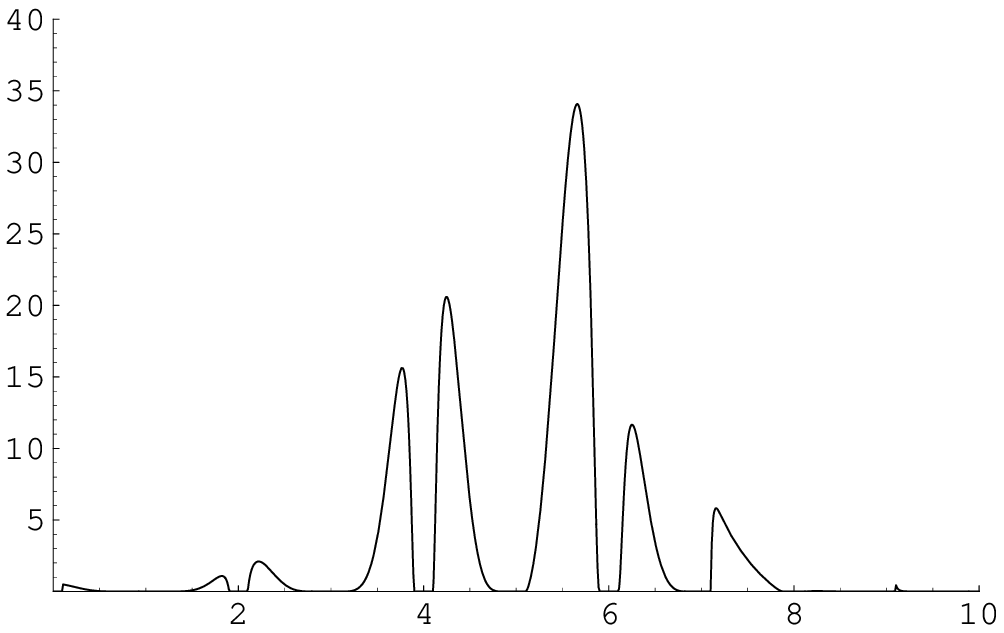, width=0.45\linewidth} 
\end{tabular}
\caption{Two example plots of the \Moo\ structure constants.
The left plot shows $c(1.5, 1.6, z)$ as a function of $z$, the right
plot shows $c(4.5, 4.6, z)$.}
\label{c=1:fig-sc}
\end{figure}

We take the fields $\phi_x$ to have the following two-- and three point
functions on the complex plane%
\footnote{We consider the relation of these correlation functions to
the operator product algebra in section \ref{sec:smear}}:
\begin{align}
  \npt{\;\phi_x(\ze_1,\zeb_1)\,\phi_y(\ze_2,\zeb_2)\;} &= 
    \delta(x-y) \cdot |\ze_{12}|^{-x^2} \;, \notag\\
  \npt{\;\phi_x(\ze_1,\zeb_1)\,
    \phi_y(\ze_2,\zeb_2)\,\phi_z(\ze_3,\zeb_3)\;} &= 
    c(x,y,z) \cdot |\ze_{12}|^{(z^2-x^2-y^2)/2} \notag\\
    & \qquad \times |\ze_{13}|^{(y^2-x^2-z^2)/2}\;
    |\ze_{23}|^{(x^2-y^2-z^2)/2} \;,
  \label{c=1:2pt3pt-bulk}
\end{align}
where $\ze_{ij}=\ze_i-\ze_j$ and
the structure constants $c(x,y,z)$ are of the form
\begin{align}
  c(x,y,z) = P(x,y,z) \cdot \exp(Q(x,y,z)) \;.
  \label{c=1:sc}
\end{align}
$P(x,y,z)$ is a step function, defined as follows:
we denote by $[x]$ the largest integer
less than
or equal to $x$ and define $f_x=x-[x]$ to be the fractional
part of $x$. Then $P(x,y,z)$ takes the form
\begin{align}
  P(x,y,z) = \begin{cases}
  \tfrac{1}{2} : &\big(\; [x]{+}[y]{+}[z] \text{ even, and } 
  |f_x{-}f_y|<f_z<\min(f_x{+}f_y\,,\,2{-}f_x{-}f_y) \;\big)\;\;\text{ or}\\
  & \big(\; [x]{+}[y]{+}[z] \text{ odd, and } 
  |f_x{-}f_y|<1{-}f_z<\min(f_x{+}f_y\,,\,2{-}f_x{-}f_y) \;\big)\\
  0 : & \text{otherwise} \end{cases}
  \label{c=1:P}
\end{align}
One can verify that $P(x,y,z)$ is symmetric in all indices and periodic
$P(x,y,z)=P(x,y,z{+}2)$. The function $Q(x,y,z)$ in \eqref{c=1:sc} is
given by 
\begin{align}
  Q(x,y,z) = \int_0^1 \hspace{-5pt}
  \frac{\D\beta}{({-}\ln\beta)\cdot(1{-}\beta)^2} \cdot \Big\{ \;
  2+\sum_{\eps=\pm 1}\big(\beta^{\eps x}{+}\beta^{\eps y}{+}
  \beta^{\eps z} \big) -
  \hspace{-10pt}\sum_{\eps_x,\eps_y,\eps_z=\pm 1} \hspace{-10pt}
  \beta^{(\eps_x x+\eps_y y+\eps_z z)/2}
  \Big\} \;. \label{c=1:Q}
\end{align}
The integral converges when $x,y,z<1$ and
$x{+}y{+}z<2$. Recall that by definition we have
$x,y,z>0$. For larger values the integral 
\eqref{c=1:Q} has to be obtained by analytic continuation in 
$x,y,z$. This makes the
formula look more cumbersome, but for numerical studies it is helpful
to have the analytic continuation done explicitly. The result is
given in appendix \ref{sec:cont}. The explicit formula also implies
that
\begin{align}
  c(x,y,z) \ge 0 \;.
\end{align}
From the formulae \eqref{c=1:P} and \eqref{c=1:Q} we see
that the structure constants $c(x,y,z)$ are symmetric in all three
indices. In figure \ref{c=1:fig-sc} the structure constants have been
plotted as functions of $z$ for two choices of the values $x,y$.
We note here that the structure constants simplify greatly if two of
$x,y$ and $z$ are half-integer, in which case $P = 1/2$ and (if $x$
and $y$ are half-integer)
$\exp(Q) = 2^{-z^2} \cdot q(z)$ where $q(z)$ is a polynomial, the
simplest case being
$c(1/2,1/2,z) = 2^{-1-z^2}$.
This is discussed in appendix \ref{sec:exactcs}.

The data given so far allows us 
(in principle) to work out all correlators on the
complex plane involving fields $\phi_x$ with $x\in\Sbb$. 
In section \eqref{sec:unprop} below we discuss consistency of the data
presented above and point out some further open questions. 

Before that we would like to define boundary states for $M_\infty$.
These enable us to work out correlators of bulk fields on the
unit disc and to compute cylinder partition functions
and check that these have the properties that we demanded in
\cite{GRW01}.


\subsection{The boundary conditions of \Moo}

\Moo\ was defined to be consistent with the boundary field theories
considered in \cite{GRW01}, so we shall first recall briefly these
results.

In a minimal model the boundary conditions are in one-one
correspondence with the conformal weights, and so can be labelled by a
pair of integers $(r,s)$. In the limit \clim\
these can all be expressed as suitable superpositions of
a fundamental set of boundary conditions $\hat a$ which are the limits
of the boundary conditions $(a,1)$.
These boundary conditions and their field content at $c=1$ are
discussed in detail in \cite{GRW01}.
Note that we do not want to imply that the set $a\in\Zbb_{>0}$ exhausts all
conformal boundary conditions of $M_\infty$. There
could well be others which can not, or only in a less direct way, 
be obtained from the minimal model limit. We have nothing to say about
those. 

The boundary conditions $\hat a$ were defined so that the Hilbert space
on the strip with boundary conditions $\hat a$ and $\hat 1$ is a
single irreducible highest-weight representation 
$L(h_{a{-}1})$ of the Virasoro
algebra of weight $h_{a{-}1}=(a-1)^2/4$. 
This in turn means that the  
partition function on the strip $Z_{\hat a \hat 1}$ is a single
Virasoro character.
If the strip is of width $\R$ and length $\L$, then 
\begin{align}
  Z_{\hat a\hat 1}(\R,\L)
= {\rm Tr}_{L(h_{a{-}1})}\left( e^{-\pi \L/\R(L_0 - c/24)} \right)
= \chi_{(a-1)}(q)
\;,
\end{align}
where 
$q = \exp(-\pi \L/\R)$ and
$\chi_x(q)$ is the character of the irreducible representations of
weight $h_x$ (given in appendix \ref{sec:repn}).
Using the modular transformation in appendix~\ref{sec:repn}, 
we can express the partition function in terms of characters of the
crossed channel (the `closed string
channel')
as
\begin{align}
  Z_{\hat a\hat 1}(\R,\L)
&= 2^{3/2}
  \int_0^\infty
  \!\!
  \sin(a\pi x)\,\sin(\pi x)\,
  \chi_{x}(\qt)\;
  \D x
\;.
\label{eq:za1}
\end{align}
where $\qt = \exp(-4\pi \R/\L)$.

Note that the degenerate representations with $x$
integer do not contribute to this partition function.
This means that we can recover \eqref{eq:za1} in the boundary state
formalism with boundary states that only include states in \Sbb.
We propose that the boundary states are given by
\begin{align}
  \is{\hat a}
= 2^{3/4}
  \int_0^\infty
  (-1)^{[x]}
  \sin(a\pi x)\,
  \ishin{x}\;
  \D x
\;.
\label{cpf:bs}
\end{align}
The factor $(-1)^{[x]}$, where $[x]$ is again the integer part
of $x$, is linked to our choice of normalisation of the bulk fields.
$\ishin{x}$ and $\ishout{x}$ are Ishibashi states \cite{Ish89}
corresponding to the bulk fields of weight $h_x$ and which satisfy
\begin{align}
  \ishout{x} \qt^{- \hf (L_0 + \bar L_0 - 1/12)} \ishin{y}
= \delta(x-y)\, \chi_x(\qt)
\;.
\end{align}

As a check we can calculate the partition function on a cylinder with
boundary conditions $\hat a$ and $\hat b$ on the two ends as
\begin{align}
  Z_{\hat a, \hat b} 
&= 
  \os{\hat a} e^{-\frac{2\pi \R}{\L}(L_0+\Lb_0-\frac{1}{12})} \is{\hat b}
= \int_0^\infty \hspace{-10pt}\D x\; 2^{3/2} \cdot
  \sin(\pi a x) \cdot \sin(\pi b x)
  \cdot \chi_x(\qt) 
= \sum_{k\in a\otimes b} \chi^{\vphantom{\phi}}_{k-1}(q) 
\;.
\label{cpf:closed}
\end{align}
Here $a\otimes b$ stands for the $su(2)$--fusion product and is a
short hand for the range of the sum
(e.g.\ for $(2)\otimes(2)=(1)+(3)$ the sum
runs over $k=1,3$).
The result \eqref{cpf:closed} agrees with that in \cite{GRW01}.

The final ingredient one needs to be able to compute all amplitudes on
the unit disk are the one--point functions of bulk fields, which one
can read off directly 
from the boundary state \eqref{cpf:bs}:
\begin{align}
  \bnpt{\phi_x(0,0)}{\hat a}{disc} = \npt{\hat a|x} = 2^{3/4} 
  (-1)^{[x]} \sin(\pi a x) 
  \;. \label{cpf:1pt}
\end{align}
To compute normalised expectation values, one needs the unit
disc partition function $Z^{\hat a}$ and the result from the minimal
model limit is
\begin{align}
  Z^{\hat a} = 2^{3/4} \pi a \;.\label{cpf:disc-pf}
\end{align}


\subsection{Unknown properties of $M_\infty$}
\label{sec:unprop}

Above we have presented expressions for correlators 
as they are obtained from the \clim\ limit of minimal
models. That the expressions are found in a limiting procedure from
well defined conformal field theories is a reason to hope, but by no
means a proof, that \Moo\ is a consistent theory by itself.

The first question is whether the correlation functions are crossing
symmetric, i.e.\ whether they are independent of how one chooses to
expand them in three point functions and intermediate states. In
section \ref{sec:cross} numerical and analytic
evidence is presented that the four point function is indeed crossing
symmetric.  

The second question is whether \Sbb\
is really the full spectrum of the theory. One check is again crossing
symmetry of the correlators, which fails if not all states are included
in the intermediate channels. On the other hand it might be possible
to add in
fields and/or states to the full theory which 
simply do not contribute as intermediate channels in the four-point
functions we consider. 

At the present stage we do not have a definite answer to this
question and it remains for future research. One may however speculate
that it is possible to extend the spectrum of fields
to all non negative real
numbers $\Rbb_{\ge0}$. As an example let us consider the possibility
to add a field of weight $h=0$ to the theory. For very small $x$, the
structure constants $c(x,y,z)$ in \eqref{c=1:sc} behave as
\begin{align}
  c(x,y,z) \sim \tfrac{1}{2} \theta(x{-}y{+}z) \theta(x{+}y{-}z)
  \qquad \; ; \; x \ll 1 \;.
  \label{c=1:c-theta-approx}
\end{align}
The function $\theta(x)$ is taken to be zero for $x\le 0$ and one for
$x>0$. Using this one finds that
$\lim_{\xlim} \npt{\;\phi_x\,\phi_y\,\phi_z\;} = \delta_{y,z}$. 
The $\delta$--symbol here is meant as a discrete (Kronecker) delta, not
as a Dirac delta distribution. So $\delta_{y,z}$ is a function 
which is one if $y=z$ and zero otherwise. In particular 
we do not recover the two point function \eqref{c=1:2pt3pt-bulk} in
this limit. Thus 
$\lim_{\xlim} \phi_x$ is not a good candidate for our
tentative field of weight zero. However, since
\begin{align}
  \tfrac{d}{dx}c(x,y,z) \sim \tfrac{1}{2} \big(
  \delta(x{-}y{+}z) + \delta(x{+}y{-}z) \big)
  \qquad \; ; \; x \ll y,z \;,
\end{align}
the field $\hat\phi_0(\ze,\zeb) = 
\lim_{\xlim} \tfrac{d}{dx}\phi_x(\ze,\zeb)$ has the property
\begin{align}
  \npt{\;\hat\phi_0\,\phi_y\,\phi_z\;} = 
  \delta(y-z) \;.
  \label{c=1:phihat-3pt}
\end{align}
We can repeat the above calculation with the alternative definition 
$\hat\phi_0(\ze,\zeb) = \lim_{\xlim} \tfrac{1}{x}\phi_x(\ze,\zeb)$
and are led to the same answer. This can be understood since
$\lim_{\xlim} \phi_x = 0$ inside correlators, at least in a
distributional sense (it can be nonvanishing on a measure zero set, as
we have seen above).
Further, using \eqref{cpf:1pt} and \eqref{cpf:disc-pf} we find that the
normalised one point function on the unit disc takes the form 
\begin{align}
  \frac{\bnpt{\hat\phi_0(0,0)}{\hat a}{disc}}{Z^{\hat a}}
  = \lim_{\xlim} \;\frac{d}{dx}\;
  \frac{ \sin(\pi a x)}{ \pi a} = 1 \;.
\end{align}
So in these two examples the field $\hat\phi_0$ behaves like an
identity field. One can now speculate that $\hat\phi_0$ is indeed a
sensible field to consider and that similar constructions
are possible for the other missing points in \Sbb.
However, our checks on crossing suggest very strongly that
these fields do not appear in the operator product expansion of the
fields in \Sbb, as we discuss in section~\ref{sec:cross}.

A third question concerns the connection of \Moo\ to Liouville
theory (see e.g.~\cite{Tes01} for a recent review). The expressions
for the bulk structure constants are very similar to the ones of
Liouville theory as given in \cite{DOt94,ZZa95}. 
In Appendix~\ref{sec:liouv}
this correspondence is made more precise. 
Furthermore, in Liouville theory one equally finds a discrete series
of boundary conditions, whose field content consist solely of
degenerate Virasoro representations \cite{ZZa01}. 
It would be interesting to
know whether \Moo\ can be obtained from Liouville theory as a
limit \clim\ from above and whether methods based on
continuous sets of representations as presented in \cite{PTe99,Tes01} could
be used to establish consistency of $M_\infty$.

A fourth point open for speculation is whether the set of conformal
boundary conditions $\hat a$, one for each positive integer, is
complete, i.e.\ if any conformal boundary condition 
of \Moo\ can be rewritten
as a superposition of the boundary conditions $\hat a$. 
Again we cannot make any definite statement, and this problem remains
open for future work. However two arguments hint that there should be
more boundary conditions. Firstly, from a RCFT point of view, there
should be one fundamental conformal boundary condition for each
diagonal field in the bulk spectrum \cite{Car89,BPPZ99}. In
\Moo\ the bulk spectrum is continuous, so one might expect a
continuum of boundary conditions. Secondly, in Liouville theory one
equally finds a continuum of conformal boundary conditions
\cite{FZZ00}. 


\subsection{Operator product expansion and the vacuum state}
\label{sec:smear}

In \eqref{c=1:2pt3pt-bulk} we have defined $M_\infty$ in the bulk via
its two and 
three point function. We would like to understand the constants
appearing in the three point function as coefficients in the operator
product expansion of two primary fields. The most obvious OPE one
could write down would be
\begin{align}
  \phi_x(\ze_1,\zeb_1)\phi_y(\ze_2,\zeb_2)
  = \int_\Sbb \!\! \D z \; c(x,y,z) \; |\ze_1-\ze_2|^{2h_z-2h_x-2h_y} \;
  \big( \phi_z(\ze_2,\zeb_2) + \text{descendants} \big) \;.
  \label{c=1:singope}
\end{align}
This expression however causes an immediate problem when inserted in
the two point function \eqref{c=1:2pt3pt-bulk}. In this case we should
recover the
$\delta$--function, but since all fields in the spectrum $\Sbb$ have
nonzero conformal weight, the RHS of the OPE 
\eqref{c=1:singope} vanishes identically.

A way around this problem can be found if one understands the states
$\is{x}$ and the corresponding fields $\phi_x(\ze,\zeb)$ in a
distributional sense. To this end let us define states and fields
with smeared out conformal weight (rather than position) as
\begin{align}
  \is{f} 
= \int_\Sbb \!\! \D x \,f(x)\, \is{x} \quad \text{ and }
  \quad \phi_f(\ze,\zeb) 
= \int_\Sbb\!\! \D x\, f(x)\, \phi_x(\ze,\zeb) \;.
\end{align}
The smeared states have the property $\npt{x|f}=f(x)$ and
the OPE of the smeared fields takes the form
\begin{align}
  \phi_f(\ze_1,\zeb_1)\phi_g(\ze_2,\zeb_2)
  = \phi_h(\ze_2,\zeb_2) + \text{descendants} \;,
  \label{c=1:smearope}
\end{align}
where
\begin{align}
  h(z) = \int_\Sbb \!\! \D x \int_\Sbb \!\! \D y \;c(x,y,z)\; f(x)\, g(y) \;
  |\ze_1-\ze_2|^{2h_z-2h_x-2h_y} \;.
  \label{c=1:ope-h-def}
\end{align}

In order to
determine the set of test functions $f$ we want to allow, let us define
a linear functional $\os{0}$ which will correspond to the out
vacuum. The expression for $\hat\phi_0$ in \eqref{c=1:phihat-3pt}
motivates the definitions 
\begin{align}
  \os{0} = \lim_{x\rightarrow 0} \tfrac{1}{x} \os{x} \;,
 \;\;\hbox{ and }\;\;
  \is{0} = \lim_{x\rightarrow 0} \tfrac{1}{x} \is{x} \;,
\label{eq:vacs}
\end{align}
As test functions we allow functions $f(x)$ such that $\is{f}$ is
in the domain of $\os{0}$, i.e.
\begin{align}
  \npt{0|f} = \lim_{x\rightarrow 0} \frac{f(x)}{x} 
  \label{c=1:<0|f>}
\end{align}
has to be well defined. This fixes the behaviour of $f$ close to
zero. Additionally we want $f$ to be square integrable and 
continuous everywhere except for jumps which happen at a finite
distance from each other (to allow for the factor $P$ in \eqref{c=1:sc}).

Similarly we can ask what properties $f(x)$ must have such that the
operator $\phi_f(\ze,\zeb)$ maps $\is{0}$ into the domain of
$\os{0}$. Using the approximation \eqref{c=1:c-theta-approx} for the
structure constants we find, for $|\ze|>0$,
\begin{align}
  \os{0}\phi_f(\ze,\zeb)\is{0} = 
  \lim_{x\rightarrow 0}  \lim_{y\rightarrow 0} 
  \frac{\os{x}\phi_f(\ze,\zeb)\is{y}}{xy}
  = \lim_{x\rightarrow 0} \frac{f(x)}{x} \;.
  \label{c=1:<0|f|0>}
\end{align}
This is the same condition as \eqref{c=1:<0|f>} for the inner product
$\npt{0|f}$, consistent with the idea that the state field
correspondence will manifest itself in the form 
$\lim_{\ze\rightarrow 0} \phi_f(\ze,\zeb)\is{0} = \is{f}$.

Let us see how the above distributional interpretation of fields and
states in $\Sbb$ resolves the problem encountered when inserting the
OPE \eqref{c=1:singope} into the two point function:
\begin{align}
  \os{0}\phi_f(\ze_1,\zeb_1)\phi_g(\ze_2,\zeb_2)\is{0} =
  \os{0}\phi_h(\ze_2,\zeb_2)\is{0} = \lim_{z\rightarrow 0} 
  \frac{h(z)}{z} = \int_\Sbb f(x) g(x)
  |\ze_1-\ze_2|^{-4h_x} \D x
  \label{c=1:two-point-ope}
\end{align}
Here we used first the definition of the OPE \eqref{c=1:smearope},
then the result for the one point function \eqref{c=1:<0|f|0>} and
finally inserted the approximation \eqref{c=1:c-theta-approx} to take the
$z\rightarrow0$ limit of \eqref{c=1:ope-h-def}. 
The original two point function in \eqref{c=1:2pt3pt-bulk} 
is exactly the formulation
of this result in terms of distributions.

The treatment above is modelled closely after a similar situation
occurring in Liouville theory. There the set of conformal weights of
states in the spectrum does not have zero as a limit point.   The
$sl(2)$--invariant vacuum is thus even at a finite distance from the
spectrum of states. As treated in detail in sections 4.6ff of
\cite{Tes01}, to define expressions like $\npt{0|f}$ in Liouville
theory, one has stronger requirements on the test function $f$, e.g.\
it should allow for an analytic continuation to zero.


\section{Crossing Symmetry}
\label{sec:cross}

The most stringent consistency condition the proposed $c=1$ theory
\Moo\  has to fulfill is that of crossing symmetry in its various
incarnations. In this section it will be verified  analytically and
numerically in two examples: for four bulk fields on the full complex
plane and for two bulk fields on the upper half plane.

In order to compute the respective correlators we will need the four
point conformal blocks, from which the correlators entering the two
examples can be constructed.


\subsection{Conformal blocks}
\label{sec:block}

To motivate the construction of conformal blocks, it is helpful to
think of a primary field $\phi_x(\ze,\zeb)$ as a linear operator on
the Hilbert space of the theory, $\phi_x(\ze,\zeb) :
\Hc\rightarrow\Hc$. The Hilbert space itself splits into
representations of the Virasoro algebra,  $\Hc = \int
L(y){\otimes}L(y) \D y$. We are deliberately vague on the meaning of
the integral, as it is not important in what follows.  Each $L(y)$ is
an irreducible Virasoro highest weight module of weight $h=y^2/4$, as
discussed in section~\ref{sec:repn}. We can now decompose the linear
map  $\phi_x(\ze,\zeb)$ into components
$V_{xy}^z(\ze){\otimes}V_{xy}^{z}(\zeb)$.  The individual  linear maps 
\begin{align}
  V_{xy}^z(\ze) : L(y) \rightarrow L(z)
\end{align}
are called {\em chiral vertex operators} (CVOs). 
In the choice of notation and presentation of the CVOs we follow
\cite{MSb89}. By construction the CVOs are required to have
the same commutation relations with the Virasoro generators as the
primary field $\phi_x(\ze,\zeb)$ itself
\begin{align}
  [L_n ,V_{xy}^z(\ze)] = \ze^n \cdot 
  \big( h_x (n{+}1) + \ze \tfrac{\D}{\D\ze} \big) V_{xy}^z(\ze) \;.
  \label{block:LV-com}
\end{align}
This defines the chiral vertex operator uniquely, up to an overall
normalisation, 
in terms of its matrix elements. Using the commutation
relations \eqref{block:LV-com}, a general matrix element
\begin{align}
  \os{z}L_{m_1} \dots L_{m_\l} \; 
  V_{xy}^z(\ze) \; 
  L_{-n_1} \dots L_{-n_k}\is{y}
\end{align}
can then be reduced
to a function of $\ze$ times $\os{z} \; V_{xy}^z(\ze) \; \is{y}$.
The latter has to be of the form 
\begin{align}
  \os{z}\;V_{xy}^z(\ze)\;\is{y} = |V_{xy}^z| \cdot \ze^{h_z-h_x-h_y} \;,
\end{align}
for some (not necessarily positive) number $|V_{xy}^z|$. This is the
remaining freedom mentioned above. It follows that for the Virasoro
algebra, the space of CVOs $V_{xy}^z(\ze)$ is at
most one dimensional. It can however have dimension zero. A simple
example of this is given by $\os{z}\;V_{x0}^z(\ze)\;\is{0}$, where the
null state $L_{-1}\is{0}$ forces 
$\tfrac{\partial}{\partial\zeta} \os{z}\;V_{x0}^z(\ze)\;\is{0} = 0$ 
and thus $|V_{x0}^z|=0$ if $x\neq z$.

If we are allowed to choose $|V_{xy}^z|$ nonzero, we fix the
normalisation of the chiral vertex operator by setting it to one.

Conformal blocks
are given in terms of products of CVOs. For example the blocks needed
for a four point correlator are
$\os{t}V_{xp}^t(1)V_{yz}^p(\ze)\is{z}$ for all possible intermediate
channels $p$. For better readability let us introduce the pictorial
notation \cite{MSb89}
\begin{align}
  \os{t}V_{xp}^t(1)V_{yz}^p(\ze)\is{z} = \bL{\cbB txyzp1\ze} \;,
  \label{block:block}
\end{align}
where $t,x,y,z,p\in\Sbb$. Note that as opposed to the situation in
minimal models, these conformal blocks do not obey any differential
equations coming from null vectors. 
They can however still be computed
order by order as a power series in $\ze$ by working out the matrix
elements of the CVOs. 

As already noted, we can think of a bulk field as a 
linear operator made up of CVOs, i.e.
\begin{align}
  \phi_x(\ze,\zeb) = \int \D y \int \D z \;
   c(x,y,z) \; V_{xy}^z(\ze)\otimes V_{xy}^z(\zeb) \;,
   \label{cross:phi-op}
\end{align}
where again we understand the integral as a formal
expression. Evaluating the bulk three point function in this
formalism, one sees that the constants $c(x,y,z)$ are indeed the bulk
structure constants appearing in \eqref{c=1:2pt3pt-bulk}. Building a
bulk four point function from the operators \eqref{cross:phi-op}, we
note that it can be expressed as a bilinear combination of conformal
blocks. 

The computation of \eqref{block:block} as a power series in
$\ze$ via explicit evaluation of the CVOs becomes very slow for higher
powers of $\ze$. 
However, there are at least
two special situations in which the chiral blocks
can be computed exactly, and an efficient algorithm due to
Al.~Zamolodchikov for numerical calculation in the remaining cases.

\subsubsection{Exact conformal blocks}
\label{sec:exactcbs}

The first special case is when \eqref{block:block} can be interpreted as a
product of free boson vertex operators, i.e.\ when the charges sum to
zero $-t{+}x{+}p{=}0$ and $-p{+}y{+}z{=}0$. For $t,x,y,z,p\in\Sbb$ 
one finds the Coulomb gas expression for the chiral four point
function 
\begin{align}
   \bL{\cbB txyzp1\ze}
=  (1{-}\ze)^{xy/2} \cdot \ze^{yz/2}
\;.
\end{align}

The second special case 
is when
\eqref{block:block} can be interpreted as a correlation function of
descendants of weight $1/16$ spin fields for a $\Zbb_2$ twisted
free boson (sometimes called the `Ramond' sector) 
\cite{Zam86,Zam87b}.
The condition for this to be the case is that
$x, y, z$ and $t$ are half integer.
The simplest example is, for $p\in\Sbb$,
\begin{align}
   \bL{{\cbB {\hf}{\hf}{\hf}{\hf}p1\ze}}
&= \frac {(16 \, q)^{h_p}}
        { [\ze(1{-}\ze)]^{1/8}\, \theta_3(q)}
 \equiv G(\ze)
\;,
\label{eq:zcb1}
\end{align}
where 
\begin{align}
 q(\ze) = \exp\big( - \frac{ \pi K(1{-}\ze) }{ K(\ze) } \big)
\;,\;\;
 K(\ze) = \frac{1}{2} \int_0^1 
 \frac{\D t}{\sqrt{ t(1{-}t)(1{-}\ze t)}}
\;,\;\;
  \theta_3(q) = \sqrt{ 2K(\ze)/\pi}
\;.
\end{align}
All the other cases can be found in terms of differential
operators acting on this function, for example the next two simplest
cases are
\begin{align}
   \bL{{\cbB {\hf}{\tf}{\tf}{\hf}p1\ze}}
&= \frac{1}{2 h_p} \, \ze^{1/2} \,
    \left[ 2 \frac{\D}{\D \ze} + \frac{1}{4 \ze(1{-}\ze)} \right]
   G(\ze)
\;,
\label{eq:zcb2}
\\
   \bL{{\cbB {\hf}{\hf}{\tf}{\tf}p1\ze}}
&= \frac{ 2}{4 h_p - 1}\,
   (1{-}\ze)^{1/2} \left[ 2 \frac{\D}{\D\ze} - \frac{1}{4 \ze(1{-}\ze)} \right]
   G(\ze)
\;.
\label{eq:zcb3}
\end{align}
For general conformal blocks, we use Zamolodchikov's two recursive
methods which are described in appendix \ref{sec:recur}.


\subsection{Four point function on the complex plane}

\begin{figure}[bt]
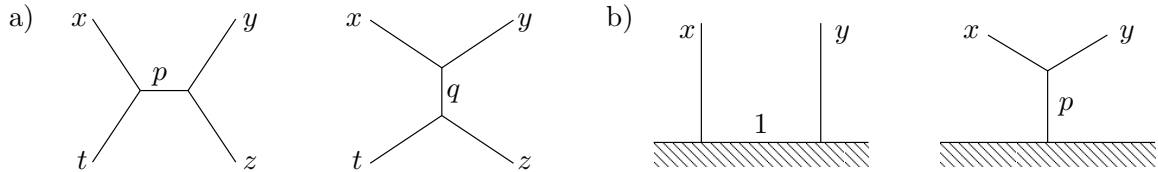

\begin{center}
\xfiga
\end{center}
\caption{Crossing symmetry of a) the four point function in the
complex plane and b) the two point function on the upper half plane.}
\label{fig:cross}
\end{figure}

We have now gathered the ingredients needed for checks of crossing
symmetry.  Let us start by considering the correlator of four bulk
fields $\phi_t$, $\phi_x$, $\phi_y$ and $\phi_z$ on the complex
plane. In figure~\ref{fig:cross}a two different ways to insert a basis
of intermediate states are depicted. In the first case one reduces the
four point function to a two point function by taking the OPEs
$\phi_t\times\phi_x$ and $\phi_y\times\phi_z$, in the second case one
considers $\phi_t\times\phi_z$ and $\phi_x\times\phi_y$. This
corresponds to two different ways of expressing the four point
function in terms of conformal blocks, as worked out below.

Using conformal transformations, we can always move three of the four
fields to the points $0,1,\infty$, say. It is thus enough to consider
the correlator
\begin{align}
  \os{t}\;\phi_x(1,1)\,\phi_y(\ze,\zeb)\;\is{z}
  = \lim_{L\rightarrow\infty} L^{4h_t}
    \npt{\,\phi_t(L,L)\,\phi_x(1,1)\,\phi_y(\ze,\zeb)\,\phi_z(0,0)\,}
  \;.
  \label{cross:4pt-by-limit}
\end{align}
We want to express
\eqref{cross:4pt-by-limit} as a bilinear combination of conformal
blocks of the form \eqref{block:block}. This corresponds to inserting
a basis of intermediate states between
$\phi_t\phi_x$ and $\phi_y\phi_z$. The coefficients in front of each 
pair of conformal blocks are given by the bulk structure constants
describing the coupling of $\phi_t\phi_x$ and $\phi_y\phi_z$ to the
intermediate channel $p$
\begin{align}
  A(\ze) &= \os{t}\;\phi_x(1,1)\,\phi_y(\ze,\zeb)\;\is{z} 
  \notag\\ &=  
  \int_\Sbb\hspace{-5pt} \D p \;\; c(t,x,p) \cdot c(y,z,p) \;
  \bL{\bL{\cbB txyzp1\ze}} \; \bL{\bL{\cbB txyzp1\zeb}} \;.
  \label{cross:4pt-A}
\end{align}
The correlator \eqref{cross:4pt-A}
has to be invariant under conformal
transformations. So if we apply for example the map 
$\xi \rightarrow 1{-}\xi$, we exchange the fields at zero and one,
$\ze$ gets mapped to $1-\ze$ while the field $\infty$ remains
fixed. In radial ordering we are now looking at the correlator 
$\os{t}\;\phi_z(1,1)\,\phi_y(1{-}\ze,1{-}\zeb)\;\is{x}$. As before we
can now insert a basis of intermediate states, this time between
$\phi_t\phi_z$ and $\phi_y\phi_x$, and arrive at the
following expression in terms of conformal blocks:
\begin{align}
  B(\ze) &= \os{t}\;\phi_z(1,1)\,\phi_y(1{-}\ze,1{-}\zeb)\;\is{x} 
  \notag\\ &=  
  \int_\Sbb\hspace{-5pt} \D q\;\; c(t,z,q) \cdot c(y,x,q) \;
  \bL{\bL{\cbB tzyxq1{1{-}\ze}}} \; \bL{\bL{\cbB tzyxq1{1{-}\zeb}}} \;.
  \label{cross:4pt-B}
\end{align}
The two function $A(\ze)$ and $B(\ze)$ now correspond to the two
different expansions in figure~\ref{fig:cross}a. Crossing symmetry
demands that they are the same $A(\ze)=B(\ze)$. 

The first case we shall consider is the correlator of four
fields of weight $h{=}\bar h{=}1/16$, since in that case both 
the bulk structure constants simplify to
$c(1/2,1/2,x) = 2^{-x^2-1}$ and we can use Zamolodchikov's exact
formula for the conformal block \eqref{eq:zcb1}.
In this case we can find the four-point function exactly
\begin{align}
  \os{\hf}\,
  \phi_{\hf}(1,1)\,
  \phi_{\hf}(\ze,\bar\ze)\,
  \is{\hf}
&= \int_\Sbb 
  c(1/2,1/2,p)^2 
  \,
   \Big|\; \bL{\bL{{\cbB {\hf}{\hf}{\hf}{\hf}p1\ze}}} \;\Big|^2
  \;\D p
\notag\\
&= \int_0^\infty 
   2^{-2 p^2 -2}
  \frac{   |\,16\, q\,|^{p^2/2}  }{  |\ze(1-\ze)|^{1/4}\, |\theta_3(q)|^2  }
  \;\D p
\notag\\
&= \frac{ \pi/2 }{4 \, |\ze(1-\ze)|^{1/4}\, 
   \big| 2 \, {\rm Re}( K(1{-}\ze)\, \overline{K(\ze)}) \big|^{1/2} }
\;.
\end{align}
Since all the fields are the same, crossing symmetry in this case is
just invariance under $\ze \to 1-\ze$, which is manifestly true.

We can also find all correlation functions of four fields with $x,y,z$
and $t$ all half-integral in the same way, 
by explicit integration
using the exact conformal block and structure constants.
For example,
the correlation function of two fields $\phi_{1/2}$ and two fields
$\phi_{3/2}$ can be decomposed in the two equivalent ways
\begin{align}
  \os{\hf}\,
  \phi_{\tf}(1,1)\,
  \phi_{\tf}(\ze,\bar\ze)\,
  \is{\hf}
&= \int_\Sbb 
  c(\hf,\tf,p)^2 
  \,
   \Big|\; \bL{\bL{{\cbB {\hf}{\tf}{\tf}{\hf}p1\ze}}} \;\Big|^2
  \;\D p
\notag\\
&= \int_\Sbb 
  c(\hf,\hf,p)\, c(\tf,\tf,p)
  \,
   \Big|\; \bL{\bL{{\cbB {\hf}{\hf}{\tf}{\tf}p1{(1{-}\ze)}}}} \;\Big|^2
  \;\D p
\;,
\end{align}
both of which integrals can be evaluated exactly and both giving the
same (less than illuminating) expression
\begin{align}
{
\pi \frac
  {
   3\,\pi^2\,|K'|^2
 - 8\,\pi\, {\rm Re}( \bar K K' ) \,{\rm Re}( E'K'  )  
 + 16\,|E'|^2\,\left({\rm Re}( \bar K K') \right)^2\,
  }
     {32
  \, \left| 1 - \ze \right|^{9/4}  
  \, |\ze|^{5/4}
  \, |K'|^2
  \, \left| 2\, {\rm Re}(\bar K K') \right| ^{5/2}  
     }
}
\;,
\end{align}
where $K\equiv K(\ze)$ and $E\equiv E(\ze)$ are the standard complete
Elliptic Integrals of the first and second kinds, $K' = K(1-\ze),
E'=E(1-\ze)$, and the bar denotes complex conjugation.

\begin{figure}[tb]
\begin{tabular}{cc}
\epsfig{file=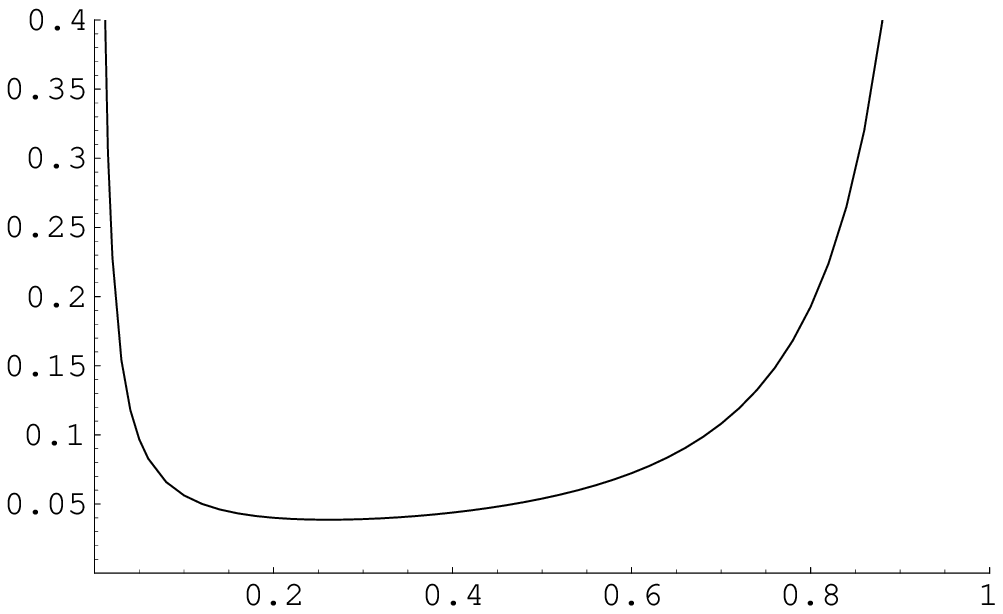, width=0.5\linewidth} &
\begin{minipage}[b]{0.4\linewidth}
\begin{center}
\begin{tabular}[b]{|l|l|l|}
\hline
$\ze$ & $A(\ze)$ & $|A(\ze){-}B(\ze)|$ \\
\hline
0.04 & 0.11802144.. & 0.00090869.. \\
0.11 & 0.05285051.. & 0.00001312.. \\
0.18 & 0.04134348.. & 0.00000041.. \\
0.32 & 0.03964800.. & 0.00000023.. \\
0.53 & 0.05830843.. & 0.00000014.. \\
0.67 & 0.09444726.. & 0.00000007.. \\
0.81 & 0.20750675.. & 0.00000136.. \\
0.88 & 0.40080739.. & 0.00003664.. \\
0.95 & 1.39461196.. & 0.00346871.. \\
\hline
\end{tabular}\end{center}
\end{minipage}
\end{tabular}
\caption{Example of crossing in the bulk with $t=0.8$,
$x=1.1$, $y=1.3$, $z=0.7$. The data has been obtained by
using the recursion relation \eqref{cross:H-recurs} to level 12 and a
numerical integration step width of $0.001$. The plot shows
the function $A(\ze)$ as given in \eqref{cross:4pt-A}
against $\ze$.}
\label{c=1:bulk-cross-data}
\end{figure}

For the remaining cases, we must use numerical tests only. 
We have used Zamolodchikov's recursive formulae for the conformal
blocks and we find excellent agreement between the 
functions $A$ and $B$. An example of 
the numerical evaluation is shown in figure~\ref{c=1:bulk-cross-data}.

\subsubsection{Crossing and degenerate representations}
\label{sec:novac}

We shall now argue that the degenerate representations, and in
particular the identity representation, do not occur as 
separate intermediate channels in the bulk four-point correlation
functions if we assume that the correlation functions are continuous
in the weights of the fields. 
For general $t,x,y,z$, we have proposed the decomposition of the
correlation function as 
\begin{align}
  \os{t}\,
  \phi_{x}(1,1)\,
  \phi_{y}(\ze,\bar\ze)\,
  \is{z}
= \int_\Sbb 
  c(t,x,p)\, c(y,z,p)
  \,
   \Big|\; \bL{\bL{{\cbB {t}{x}{y}{z}p1\ze}}} \;\Big|^2
  \;\D p
\;.
\label{eq:new1}
\end{align}
The degenerate representations with $p$ integer all contain singular
vectors, and so the chiral blocks in which they can occur are limited 
by the requirement that these singular vectors decouple -- for example,
the `vacuum' channel $p=0$ is only allowed if $t=x$ and $y=z$.

If $t,x,y$ and $z$ satisfy such requirements, then we could 
{\em a priori} consider the more general expression
\begin{align}
  \os{t}\,
  \phi_{x}(1,1)\,
  \phi_{y}(\ze,\bar\ze)\,
  \is{z}
&=
  \int_\Sbb 
  c(t,x,p)\, c(y,z,p)
  \,
   \Big|\; \bL{\bL{{\cbB {t}{x}{y}{z}p1\ze}}} \;\Big|^2
  \;\D p
\notag\\
&+
  \sum_{p=0}^\infty
  \tilde c(t,x,p)\, \tilde c(y,z,p)
  \,
   \Big|\; \bL{\bL{{\cbB {t}{x}{y}{z}p1\ze}}} \;\Big|^2
\;.
\label{eq:new2}
\end{align}
However, provided the integrand remains finite as
$t,x,y$ and $z$ approach values for which degenerate intermediate
channels are allowed, the integral as a whole is continuous 
(and remains crossing symmetric), and so continuity of the whole correlation
function implies that the couplings to the degenerate blocks $\tilde
c(t,x,p)$ are all zero.
In particular the `vacuum' channel $p=0$ never occurs as an intermediate
channel.

We note here that we have not proven that the integrand remains
finite, and indeed there are limiting values of $t,x,y,z$ for which
the limits of the chiral blocks develop poles, but in all cases we
have investigated these poles are cancelled by zeros in the structure
constants.


\subsection{Two point function on the upper half plane}

Rather than working with the unit disc, it is convenient to use a
conformal mapping to the upper half plane. On the UHP we can use
directly a method introduced by Cardy \cite{Car84} to express the two
point function in terms of chiral four point blocks
\eqref{block:block}. 

Since in section \ref{sec:c=1} we have only presented the one point
function \eqref{cpf:1pt} in the presence of a boundary, 
we can only say how a bulk
field couples to the identity on the boundary. To describe the
coupling to other boundary fields we would need correlators involving
one bulk and one boundary field. 

In order to check crossing symmetry in presence of a boundary we have
to restrict ourselves for the time being to boundary conditions on
which the unique primary boundary field is the identity. Among the set
of boundary conditions presented in section \ref{sec:c=1} there is
exactly one fulfilling the criterion, the boundary condition 
$\hat 1$.

The statement of crossing symmetry for the two point function on the
UHP is depicted in figure~\ref{fig:cross}b. On the one hand we can
expand both bulk fields in terms of boundary fields, i.e.\ only the
identity in our case. On the other hand we can compute the OPE in the
bulk and evaluate the one point function of the resulting field.

Denote as $\B{\hat a}{x}{\Id}$ the constant describing the coupling of
a bulk field to the identity on the $\hat a$--boundary, i.e.\
\begin{align}
  \frac{\bnpt{\phi_x(\ze,\zeb)}{\hat a}{disc}}{Z^{\hat a}} =
  \B{\hat a}{x}{\Id} \cdot ( 1-|\ze|^2)^{-2h_x} \qquad
  ,\quad
  \B{\hat a}{x}{\Id} = (-1)^{[x]} \cdot
  \frac{\sin(\pi a x)}{\pi a} \;,
\end{align}
where the numerical value of $\B{\hat a}{x}{\Id}$ follows from 
\eqref{cpf:1pt} and \eqref{cpf:disc-pf}. Since unit disc and UHP are
related by a conformal transformation one sees that the values of the
bulk boundary couplings are the same in both cases.

Coupling both bulk fields to the identity on the boundary we find the
correlator 
\begin{align}
  A(a,b) &= \frac{\,\bnpt{\phi_x(a{+}ib)\,\phi_y(ib)\,}{\hat 1}{UHP}}{Z} 
  \notag\\ &=
  (2b)^{-2h_x-2h_y} \, \eta^{2h_y} \cdot 
  \B{\hat 1}{x}{\Id} \, \B{\hat 1}{y}{\Id} \cdot 
  \bL{\bL{\cbB xxyy{\Id}1{\eta}}}
  \qquad , \; \eta = \frac{4b^2}{a^2+4b^2} \;.
  \label{cross:2pt-A}
\end{align}
Taking the OPE of the two bulk fields first we obtain
\begin{align}
  B(a,b) &= \frac{\bnpt{\,\phi_x(a{+}ib)\,\phi_y(ib)\,}{\hat 1}{UHP}}{Z} 
  \notag\\ &=
  (2b)^{-2h_x-2h_y} \, \eta^{2h_y} \cdot
  \int_\Sbb\hspace{-3pt}\D p \;\; c(x,y,p) \,
  \B{\hat 1}{p}{\Id} \cdot \bL{\bL{\cbB xyyxp1{1{-}\eta}}} \;,
  \label{cross:2pt-B}
\end{align}
where $\eta$ is the same as in equation \eqref{cross:2pt-A}. Note that
the conformal block in \eqref{cross:2pt-A} involves a degenerate
representation in the intermediate channel, while this is not the case
for the blocks appearing in \eqref{cross:2pt-B}.

\begin{figure}[tb]
\begin{tabular}{cc}
\epsfig{file=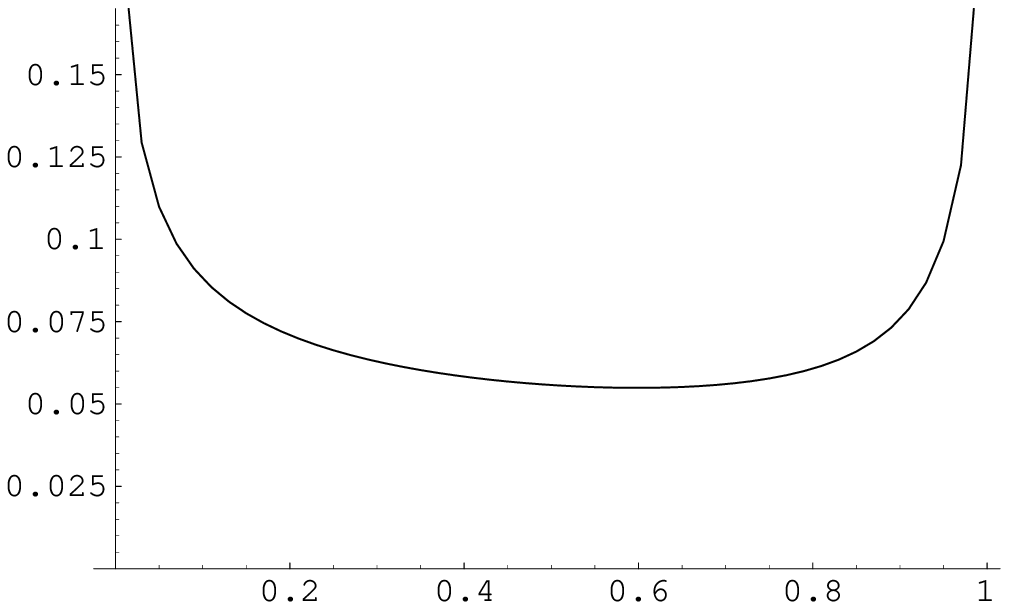, width=0.5\linewidth} &
\begin{minipage}[b]{0.4\linewidth}
\begin{center}
\begin{tabular}[b]{|l|l|l|}
\hline
$\ze$ & $A_b(\ze)$ & $|A_b(\ze){-}B_b(\ze)|$ \\
\hline
0.04 & 0.11798728.. & 0.00003262.. \\
0.11 & 0.08548602.. & 0.00000313.. \\
0.18 & 0.07325737.. & 0.00000054.. \\
0.32 & 0.06177076.. & 0.00000000.. \\
0.53 & 0.05532286.. & 0.00000022.. \\
0.67 & 0.05543060.. & 0.00000069.. \\
0.81 & 0.06154874.. & 0.00002715.. \\
0.88 & 0.07096936.. & 0.00017047.. \\
0.95 & 0.09943759.. & 0.00198972.. \\
\hline
\end{tabular}\end{center}
\end{minipage}
\end{tabular}
\caption{Example of crossing on the upper half plane with
$x=1.25$ and $y=0.8$. The data has been obtained by
using the $c$--recursion relation \eqref{cross:H=0-block} for internal
channel zero to level~8. In the crossed channel the $H$--recursion
\eqref{cross:H-recurs} 
was used to level~6 together with a 
numerical integration step width of $0.005$. The plot shows
the function $A_b(\ze)$ as given in \eqref{cross:AbBb-def}
against $\ze$.}
\label{c=1:bnd-cross-data}
\end{figure}

One can now verify crossing symmetry by checking that
$A(a,b)=B(a,b)$ as a function of the two real parameters $a,b$. 
Instead of looking at $A$ and $B$, we can consider the functions
\begin{align}
  A_b(\eta)=(2b)^{2h_x+2h_y} \eta^{-2h_y} A(a,b)
  \quad \text{ and } \quad
  B_b(\eta)=(2b)^{2h_x+2h_y} \eta^{-2h_y} B(a,b) \;,
  \label{cross:AbBb-def}
\end{align}
which only depend
on the crossing ratio $\eta$. 

We again find excellent numerical 
agreement between $A_b(\eta)$ and $B_b(\eta)$, an example is given in 
figure \ref{c=1:bnd-cross-data}.


\section{The limit of Minimal Models}
\label{sec:lim}

This more technical section treats in some detail 
how we define the $c=1$ theory 
\begin{align}
  M_\infty = \lim_{\plim} M_p \;.  
\end{align}
In particular it is shown how the expressions \eqref{c=1:2pt3pt-bulk} and
\eqref{c=1:sc} can be obtained. The contents of this section should be
understood as a motivation for 
\eqref{c=1:2pt3pt-bulk}, \eqref{c=1:sc} and
does not constitute a proof.

As a first step it is shown that the minimal model bulk spectrum
approaches a continuum as \clim. Then the $M_\infty$
fields are defined as an average over minimal model bulk fields
of approximately the same weight. This prescription is used to compute
the correlators of $M_\infty$.


\subsection{Limit of the spectrum}
\label{sec:lots}

Consider the unitary minimal model $M_p$. With $t=\tfrac{p}{p+1}$ and
$\Delta=1{-}t=\tfrac{1}{p+1}$ its central charge is given by
\begin{align}
  c \;=\; 13 - 6(t{+}t^{-1}) \;=\; 1 - 6\Delta^2 + O(\Delta^3) \;<\; 1
\end{align}
The limit \clim\ corresponds to $\Delta\rightarrow 0$ or
equivalently to \plim.

The bulk field content of the
diagonal modular invariant is given by the Kac-table: There is one bulk field
$\phi_{r,s}(\ze,\zeb)$ for each pair of integers $r,s$ 
with $1\le r \le p{-}1$ and 
$1\le s\le p$, subject to the identification $(r,s)\sim(p{-}r,p{+}1{-}s)$. 
With $d_{ab} = a - bt$ the
left/right conformal weight of $\phi_{r,s}$ is given by
\begin{align}
  h_{r,s} = h_{p{-}r,p{+}1{-}s} = 
  \tfrac{1}{4t}\big(\,{d_{rs}}^2 - {d_{11}}^2\,\big)
  \;. \label{lim:hrs}
\end{align}
Taking $\Delta\rightarrow 0$, each individual weight approaches
the limiting value
\begin{align}
  h_{r,s} \xrightarrow{\quad\Delta\rightarrow 0\quad}
  (r{-}s)^2 / 4 \;.
\end{align}
However, while each weight approaches 
one of these discrete values, they do so
in such a way that the spectrum as a whole approaches a continuum.

To see this, we first pick a range $K$ 
of Kac-labels such that each field is
represented once. It will be convenient 
to choose the subset above the
diagonal, that is we choose the set
\begin{align}
  K = \big\{\;(r,s) \;\big|\; 1\,{\le}\,r\,{\le}\,p{-}1\,,\;
  1\,{\le}\,s\,{\le}\,p\,,\; d_{rs}{>}0 \;\big\} \;.
  \label{lim:Kset}
\end{align}
Recall that the \Moo\ fields $\phi_x(\ze,\zeb)$ were labelled by
a real number $x$, which was linked to the conformal weight via
$h_x=x^2/4$. 
Let $x_{rs} = 2 \sqrt{h_{rs}}$. 
The set $K$ can be split into slices which contain
fields with approximately the same weight. We will parametrise these slices
by the \Moo\ label $x$: denote as $N(x,\eps)$ the subset of $K$
which contains all labels $(r,s)$ with $x_{rs}$ in the interval
$[x,x{+}\eps[$,  
\begin{align}
  N(x,\eps) = \big\{\; (r,s)\in K \;\big|\; 
  x \le x_{rs} < x{+}\eps \; \big\} \;.
\end{align}
We will need a more direct way to describe the Kac labels 
$(r,s)\in N(x,\eps)$. Working to first order in $\Delta$, we find
\begin{align}
  x_{rs} = r{-}s + \Delta\tfrac{r+s}{2} + O(\Delta^2) \;.
  \label{lim:xrs-1st-order}
\end{align}
From the possible ranges of $r,s$, i.e.\ $1\le r \le p{-}1$ and
$1\le s \le p$, it follows that
$0<\Delta\tfrac{r{+}s}{2}<1$. Denoting 
again by $[x]$ the largest integer
less or equal to $x$, we can thus write
\begin{align}
  x_{rs} = [x_{rs}] + f_{rs}
  \quad \text{ where }\quad 
  [x_{rs}] = r-s \; , \quad
  f_{rs} = \Delta\cdot s + \Delta \tfrac{[x_{rs}]}{2} \;.
  \label{lim:xrs-frs}
\end{align}
Looking again at the allowed ranges of $r,s$ we find that the range of
$f_{rs}$ is restricted to
\begin{align}
  \Delta\cdot \tfrac{2+[x_{rs}]}{2} \;\le\; f_{rs} 
  \;\le\; 
  1 - \Delta \cdot \tfrac{4+[x_{rs}]}{2}  \;.
\end{align}
For fixed $\Delta$ there is thus an interval around each integer 
which is not accessible in the given minimal model. Outside this 
interval \eqref{lim:xrs-frs} shows that the values of $x_{rs}$ are
distributed uniformly (up to $O(\Delta^2)$) with separation $\Delta$. 

The above shows that for $x\notin\Zbb$ and for $\Delta,\eps$ small
enough, the following more direct description of the set $N(x,\eps)$ is
possible:
\begin{align}
  N(x,\eps) \;=\; \big\{\; (\,[x]{+}n,n\,) \in K \;\big|\; 
  n{\in}\Zbb \text{ and }
  \tfrac{f_x}{\Delta}{+}\tfrac{[x]}{2} \, \le \,n\,<\,
  \tfrac{f_x+\eps}{\Delta}{+}\tfrac{[x]}{2} \; \big\} \;.
  \label{lim:Nnew}
\end{align}
Here $f_x$ denotes again the fractional part of $x$, 
i.e.\ $f_x=x-[x]$. In other words, the set $N(x,\eps)$ thus simply
consists of Kac labels $(r,s)$ where $s$ takes all values in a certain
range and the difference $r{-}s$ is fixed to be $[x]$. 

In particular \eqref{lim:Nnew} shows that the number of elements in
each set $N(x,\eps)$ is given by
\begin{align}
  |N(x,\eps)| \sim  \eps / \Delta \;.
\end{align}
For $\Delta$ small enough, we can find arbitrarily many
labels in the set $N(x,\eps)$, that is arbitrarily many bulk fields 
$\phi_{r,s}$ have a conformal weight $h_{r,s}$ close to
$h_x$. Furthermore the leading behaviour of $|N(x,\eps)|$ is 
independent of $x$. Thus
the spectrum of primary fields approaches a continuum
with a uniform spectral density, when parametrised by $x$. Had we
chosen the conformal weight $h$ as the parameter instead of $x$, the
spectral density would not be constant and less convenient to work with.


\subsection{Normalisation of $M_p$}
\label{sec:Mp-norm}

In order to get finite expressions for the quantities of our interest
in the limit $\Delta\rightarrow 0$, we have to normalise
the individual unitary minimal models $M_p$ in an appropriate way. 

To do so we first choose a particular solution for the minimal model
structure constants. Then free parameters are inserted, which reflect
the freedom to redefine the fields. 

As already mentioned in section \ref{sec:lots},
the minimal model $M_p$ has one bulk field $\phi_i(\ze,\zeb)$ for each
pair of Kac labels $i\in K$, where $K$ is defined as in 
\eqref{lim:Kset}. For these we need to specify the bulk structure
constants $\C ijk$ that appear in the OPE of two primary fields
\begin{align}
  \phi_i(\ze,\zeb)\phi_j(0,0) =
  \sum_k \C ijk |\ze|^{2h_k-2h_i-2h_j} (\phi_k(0,0) +
  \text{ descendant fields } ) \;.
\end{align}

We are also interested in correlation functions in the unit disc. To
compute these we first need to know the set of conformal boundary
conditions for the model $M_p$.

We will distinguish between
fundamental boundary conditions and superpositions. Fundamental
boundary conditions can not be expressed as a sum (with positive integer
coefficients) of other conformal boundary conditions. All
superpositions on the other hand are sums of fundamental boundary
conditions. In terms of boundary fields, fundamental boundary
conditions are identified by the property that exactly one field of
conformal weight zero lives on them.

The fundamental conformal boundary conditions 
of $M_p$ have been found by Cardy
\cite{Car89}. There is a finite number of them, and they are in one to one 
correspondence with the Kac labels in $K$. So we can label boundary
conditions by $a\in K$. 

The correlators on the unit disc with boundary condition $a$ are now
specified once we know the 
(unnormalised) one point functions of the primary bulk
fields \bnpt{\phi_i(0,0)}{a}{disc} and the unit disc partition
function $Z^a$.

The solution for $M_p$ 
that serves as a starting point uses the bulk structure
constants $\C ijk$ 
by Dotsenko and Fateev, discussed below in section~\ref{sec:mm-bulk-sc}
and the unit disc expectation values given by Cardy and Lewellen
\cite{CLe91}. 
In our normalisation these
have the properties
\begin{align}
  \npt{0|0} = 1 \;,\quad
  \C ii1 = 1 \;,\quad
  Z^a = \frac{\S{\Id}{a}}{\sqrt{\S{\Id}{\Id}}} \;,\quad
  \bnpt{\phi_i(0,0)}{a}{disc} = 
  \sqrt{\frac{\S{\Id}{\Id}}{\S{i}{\Id}}} \cdot
  \frac{\S{i}{a}}{\S{\Id}{a}} \cdot Z^a \;.
  \label{lim:Mp-fixed}
\end{align}
Here $\S ij$ denotes the modular $S$--matrix that appears in the
transformation of characters. For minimal models it is given by
\begin{align}
  \S{rs}{r'\!s'} = 2^{3/2} \big(p(p{+}1)\big)^{-1/2} (-1)^{1+rs'+sr'}
  \cdot \sin(\pi rr'/t) \cdot \sin(\pi ss' t) \;.
  \label{lim:s-mat}
\end{align}

The solution \eqref{lim:Mp-fixed} is 
in fact unique, up to the freedom to rescale the
fields and a choice of normalisation for the bulk vacuum,
\begin{align}
  \phi_i(\ze,\zeb) \rightarrow \alpha_i \cdot \phi_i(\ze,\zeb) 
  \quad,\qquad \is{0} \rightarrow \gamma \cdot \is{0} \;.
  \label{lim:fieldredef}
\end{align}
Implementing this choice of normalisation explicitly we can rewrite
\eqref{lim:Mp-fixed} as
\begin{align}
  \npt{0|0} = \gamma(\!\Delta\!)^2 \;,\quad
  \C ii1 = \alpha_i(\!\Delta\!)^2 \;,\quad
  \C ijk = \frac{\alpha_i(\!\Delta\!)\alpha_j(\!\Delta\!)}{\alpha_k(\!\Delta\!)}
     \cdot {\tilde C}_{ij}^{\;k} \;,\notag\\
  Z^a = \gamma(\!\Delta\!) \cdot \S{\Id}{a} / \sqrt{\S{\Id}{\Id}} \;,\quad
  \bnpt{\phi_i(0,0)}{a}{disc} = \alpha_i(\!\Delta\!) \cdot
  \sqrt{\S{\Id}{\Id} / \S{i}{\Id}} \cdot
  \S{i}{a} / \S{\Id}{a} \cdot Z^a \;.
  \label{lim:Mp-var}
\end{align}
Here ${\tilde C}_{ij}^{\;k}$ refers to the bulk structure constants in the
normalisation \eqref{lim:Mp-fixed}, i.e.\ ${\tilde C}_{ii}^{\;1}=1$.

One may wonder how the normalisation of the bulk vacuum enters the
unit disc amplitude $Z^a$. This is due to the boundary state
formalism. If we describe the boundary of the unit disc as an out
state $\os{a}$, then we would like it to reproduce the unit disc
expectation values given above
\begin{align}
  \bnpt{\phi_i(0,0)}{a}{disc} = \npt{a|i} \quad,\quad
  Z^{a} = \npt{a|0} \;.
  \label{lim:unit-bs}
\end{align}
Furthermore, the normalisation of the boundary state $\os{a}$ is
constrained by the cylinder partition function. Consider a cylinder of
length $R$ and circumference $L$ with conformal boundary conditions
$a,b$. Its partition function in the open
and closed string channel is given by, respectively, \cite{Car89}
\begin{align}
  Z_{\text{cyl}} = \os{a}e^{\frac{-2\pi R}{L}(L_0+\bar L_0
  -\frac{c}{12})}\is{b} = \sum_i {N_{ia}}^b \chi_i(e^{-\pi\,L/R}) \;.
\end{align}
Here $\chi_i$ denotes the character of the Virasoro highest weight
representation of weight $h_i$ and ${N_{ij}}^k$ are the Verlinde fusion
numbers. Note that the r.h.s. of this expression is independent of any
normalisation. If we take the large $R$ limit on the lhs, only the
ground state will contribute and we get
\begin{align}
  Z_{\text{cyl}} \sim \frac{\npt{a|0}\npt{0|b}}{\npt{0|0}}
  \cdot e^{-R E_0(L)} \;,
  \label{lim:cyl-lim}
\end{align}
where $E_0(L)=-\tfrac{c}{6L}$.
Requiring \eqref{lim:cyl-lim} to be independent of the choice of $\gamma$ in
\eqref{lim:fieldredef}, together with \eqref{lim:unit-bs}, leads to 
the $\gamma$--dependence as in \eqref{lim:Mp-var}.

The parameters $\alpha_i(\Delta)$ and $\gamma(\Delta)$ introduced in
the solution \eqref{lim:Mp-var} will be fixed in the next section to
obtain the desired behaviour of correlators in the
$\Delta\rightarrow 0$ limit.


\subsection{Correlators of averaged fields}

The basic idea in taking the $\Delta\rightarrow 0$ limit in the bulk
is that the fundamental parameters of minimal models  are not the Kac
labels $(r,s)$, but the conformal weights $h_{rs}$.  So rather than
taking the limit for a fixed pair of Kac labels $(r,s)$ we take it for
a fixed value of $h$. This means that the pairs $(r,s)$ now depend on
$\Delta$ in such a way that $h_{rs}\rightarrow h$ as
$\Delta\rightarrow 0$.

In fact, choosing a specific sequence of Kac labels to approach a
given weight $h$ may cause problems. Consider for example the
sequences $i_p=\big(2[\tfrac p5],2[\tfrac p5]\big)$ and
$j_p=\big(2[\tfrac p5]+1,2[\tfrac p5]+1\big)$. Here the notation $[x]$
refers again to the largest integer $\le x$.  Both, $i_p$ and $j_p$
are Kac labels in the minimal model $M_p$. Both sequences approach the
limiting weight $h=1/25$. But the fusion  $i_p \times i_p \rightarrow
i_p$ is forbidden, while the fusion $j_p \times j_p \rightarrow j_p$
is allowed. So taking the limit of the bulk structure constants $\C
iii$ will automatically give zero, while the limit of $\C jjj$ is
potentially nonzero.

To avoid this sort of difficulty one can consider averages over
primary fields. That is, instead of taking limits of individual
fields, in each model $M_p$ we sum over all fields whose weight lies
in a certain neighbourhood of $h$.

For the computation below it is preferable to use the parameter
$x=2\sqrt{h}$ instead of $h$ itself, since as we have seen in section
\ref{sec:lots} the spectrum then approaches the continuum with a
uniform density. 
The averaged fields $V_{x,\eps}(\ze,\zeb)$ we will consider are
defined as sums over primary fields with Kac labels in the set 
$N(x,\eps)$:
\begin{align}
  V_{x,\eps}(\ze,\zeb) = A_x(\Delta,\eps)
  \cdot \sum_{rs\in N(x,\eps)} \phi_{r,s}(\ze,\zeb)
  \label{lim:Vdef}
\end{align}
The prefactor $A_x(\Delta,\eps)$, together with
$\alpha_i(\Delta)$ and $\gamma(\Delta)$ from section
\ref{sec:Mp-norm} will be adjusted later to get finite
answers for the correlators. It will turn out that $A_x$ can be chosen
to be independent of $x$ and $\alpha_i$ independent of $i$. This means
we can restrict the set of possible gauge choices to 
$A_x(\Delta,\eps)=A(\Delta,\eps)$ and
$\alpha_i(\Delta)=\alpha(\Delta)$, for some $A, \alpha$, and still
obtain finite answers for the $n$--point functions.

To obtain the \Moo\ fields $\phi_x(\ze,\zeb)$ defined in section
\ref{sec:c=1}, one starts from the averaged fields
$V_{x,\eps}(\ze,\zeb)$ and first takes the limit $\Delta\rightarrow 0$,
and then $\eps\rightarrow 0$
\begin{align}
  \phi_x(\ze,\zeb) = \lim_{\eps\rightarrow 0} 
  \lim_{\Delta\rightarrow 0} V_{x,\eps}(\ze,\zeb) \;.
  \label{lim:avfield}
\end{align} 
Note that it does not make sense to exchange
the limits, since at finite $\Delta$ only finitely many primary fields
are present in the model, and taking $\eps$ to zero would in
general leave no fields in the interval $[x,x{+}\eps]$ to be averaged
over.

For finite $\Delta$ we would expect the averaged fields
$V_{x,\eps}$ to be in some sense `close' to the analogous average
over fields at $\Delta=0$,
\begin{align}
  V_{x,\eps}(\ze,\zeb) \approx \frac{1}{\eps} 
  \int_x^{x+\eps} \hspace{-10pt}\D x' \; \phi_{x'}(\ze,\zeb) \;.
  \label{lim:Vapprox}
\end{align}
This is the prescription we will use to compute the various 
correlators below.

\subsubsection{Bulk two point function}

For the bulk two point function in \Moo\ we would like the
property 
\begin{align}
  \npt{\,\phi_x(\ze,\zeb)\,\phi_y(0,0)\,} = \delta(x{-}y) \cdot |\ze|^{-4h_x} \;.
  \label{lim:two-pt-c=1}
\end{align}
From here on we suppress the coordinate dependence of
the correlators for clarity. The letters $x,y,z$ always label fields,
and never coordinates.
In terms of \eqref{lim:Vapprox} the desired behaviour for the two
point function implies that
\begin{align}
  \frac{1}{\eps^2} 
  \int_x^{x+\eps} \hspace{-10pt}\D x'
  \int_y^{y+\eps} \hspace{-10pt}\D y' \; 
  \npt{\,\phi_{x'}\,\phi_{y'}\,} \;=\; \frac{\eps-|x{-}y|}{\eps^2} 
    \cdot \theta(\eps-|x{-}y|) \;.
\end{align}
Inserting the definition \eqref{lim:Vdef} we find that for 
minimal models
\begin{align}
  \npt{\,V_{x,\eps} \, V_{y,\eps}\,} \;&=
  \sum_{i\in N(x,\eps)}\sum_{j\in N(y,\eps)}
  A(\Delta,\eps)^2\cdot
  \npt{\,\phi_i\, \phi_j\,} \;= \hspace{-5pt}
  \sum_{i \in N(x,\eps) \cap N(y,\eps)} \hspace{-10pt}
  A^2 \cdot \alpha(\Delta)^2 \cdot \gamma(\Delta)^2 
  \notag\\
  &\approx A^2 \cdot \alpha^2 \cdot \gamma^2
  \cdot \frac{\eps-|x{-}y|}{\Delta} \cdot \theta(\eps-|x{-}y|) \;.
\end{align}
To obtain the desired behaviour \eqref{lim:two-pt-c=1} we thus need
\begin{align}
  A^2 \cdot \alpha^2 \cdot \gamma^2 = \frac{\Delta}{\eps^2} \;.
  \label{lim:bulk-2pt-cond}
\end{align}

\subsubsection{Unit disc one point function}

As already mentioned in section \ref{sec:Mp-norm} the set of distinct
fundamental conformal boundary
conditions is given by $K$ as defined in \eqref{lim:Kset}. 
Using the
results of \cite{RRS00} it was argued in \cite{GRW01} that in the
limit \clim\ not all boundary conditions remain
independent. Instead one has the formal relations
\begin{align}
  \lim_{\clim} \; (r,s) \; = \;
  \lim_{\clim} \sum_{a\in r\otimes s} (a,1) \;.
  \label{lim:bc-sup}
\end{align}
Here $a{\in}r{\otimes}s$ is the short hand notation introduced in
equation \eqref{cpf:closed}. 

In words the meaning of \eqref{lim:bc-sup} is that in the 
\clim\ limit a general boundary condition $(r,s)$
becomes indistinguishable form a superposition of boundary conditions
of the type $(a,1)$ and is thus no longer a fundamental boundary
condition itself. 

To find all boundary conditions in \Moo\ that appear in this limit
of minimal models, it is thus enough to consider the
$(a,1)$--boundary conditions. We will denote the \clim\
limit of $(a,1)$ as $\hat a$.

To obtain the limit of the boundary one point functions we can
directly use the definition \eqref{lim:avfield}
\begin{align}
  \bnpt{\,\phi_x(\ze,\zeb)\,}{\hat a}{disc} = 
  \lim_{\eps\rightarrow 0} 
  \lim_{\Delta\rightarrow 0}\; 
  \bnpt{\,V_{x,\eps}(\ze,\zeb)\,}{(a,1)}{disc} = 
  \lim_{\eps,\Delta} 
  \sum_{i\in N(x,\eps)} A \cdot \bnpt{\,\phi_i(\ze,\zeb)\,}{(a,1)}{disc}\;.
  \label{lim:one-point-average}
\end{align}
In order to compute the average using the expressions in 
\eqref{lim:Mp-var} we need to consider two different
limits of the $S$--matrix \eqref{lim:s-mat}. In the first limit we keep
the Kac labels fixed and take $\Delta$ to zero. The leading behaviour
of $\S{11}{a1}$ in this case can be found directly from
\eqref{lim:s-mat} to be
\begin{align}
  \S{11}{a1} \sim 2^{3/2} \pi^2 \Delta^3 \cdot a \;.
  \label{lim:s-mat-lim1}
\end{align}
In the second limit we consider the leading behaviour as $\Delta$ goes
to zero of $\S{rs}{a1}$ where $r,s$ varies with $\Delta$ such that
$x_{rs}=x$ is kept (approximately) fixed. To compute this limit recall
from \eqref{lim:xrs-1st-order} that $x_{rs} = r-st+\Delta\cdot[x]/2$ and note that
\eqref{lim:s-mat} can be rewritten in the form
\begin{align}
  \S{rs}{jk} = 2^{3/2} \big(p(p{+}1)\big)^{-1/2}
  \cdot \sin(\pi j (r{-}st)/t) \cdot \sin(\pi k (r{-}st) ) \;.
  \label{lim:s-mat-rs}
\end{align}
The leading behaviour of $\S{rs}{a1}$ with $x_{rs}=x$ kept fixed is
given by
\begin{align}
  \S{rs}{a1} \sim 2^{3/2} \Delta \cdot \sin( \pi x ) \sin( \pi a x) \;.
  \label{lim:s-mat-lim2}
\end{align}
Using \eqref{lim:s-mat-lim1} and \eqref{lim:s-mat-lim2} 
we can compute the leading behaviour of the unit disc partition
function \eqref{lim:Mp-fixed} and the averaged 
one point function \eqref{lim:one-point-average} to be
\begin{align}
  Z^{\hat a} &\sim 2^{3/4} \, \pi \, a \cdot \gamma \cdot \Delta^{3/2} \;,
  \notag\\
  \bnpt{\phi_x}{\hat a}{disc} &\sim  
  2^{3/4} \, (-1)^{[x]} \, \sin(\pi a x) \cdot A \, \alpha \,\gamma \cdot 
  \Delta^{-1/2} \, \eps \;.
\end{align}
If we want $\bnpt{\phi_x(0,0)}{\hat a}{disc}$ to remain finite in the
limit we have to require
\begin{align}
  A\,\alpha\,\gamma \sim \text{(const)}\cdot \frac{\Delta^{1/2}}{\eps} \;.
  \label{lim:unit-1pt-cond}
\end{align}

\subsubsection{Bulk three point functions}

To find the bulk correlator of three \Moo\ fields we use again
the prescription \eqref{lim:avfield}.  For clarity we drop the coordinate
dependence in the following formulae for the OPEs, the parameters
$x,y,z$ refer to \Moo\ field labels:
\begin{align}
  \npt{\,\phi_x\,\phi_y\,\phi_z} =   
  \lim_{\eps\rightarrow 0} 
  \lim_{\Delta\rightarrow 0} 
  \sum_{i,j,k} A^3 \C jki \C ii1 \npt{0|0} =
  \lim_{\eps,\Delta} 
  A^3 \alpha^3 \gamma^2 \sum_{i,j,k} {\tilde C}_{ij}^{\;k}
  \;,
  \label{lim:bulk-3pt-lim}
\end{align}
where the sum runs over $i\in N(x,\eps)$, $j\in N(y,\eps)$
and $k\in N(z,\eps)$. 

To compute the leading behaviour of the sum 
\eqref{lim:bulk-3pt-lim}, the most important input is an observation
due to Dotsenko \cite{Dp.c.}: The normalised structure constants
${\tilde C}_{ij}^{\;k}$ are continuous in the parameters
$d_i=r_i-s_i t$, $d_j=r_j-s_j t$, $d_k=r_k-s_k t$. 
That is, if the index combination is allowed by the fusion rules,
structure constants with similar $d_{i}, d_{j}, d_{k}$ will have a
similar value. Furthermore they do stay finite in the limit
$\Delta\rightarrow 0$.

So, as was the case for the one point functions on the unit disc,
to work out the sum in \eqref{lim:bulk-3pt-lim}, we can treat 
${\tilde C}_{ij}^{\;k}$ as independent of $i,j,k$, 
provided that we can estimate
how many triples in the set 
$N(x,\eps)\times N(y,\eps)\times N(z,\eps)$ are
allowed by fusion. In section \ref{sec:fusion} we will argue that
in the limit
$\eps,\Delta\rightarrow0$, depending on the choice of $x,y,z$
either no triples are allowed or exactly
half of them.  

This gives the leading behaviour
\begin{align}
  \npt{\,\phi_x\,\phi_y\,\phi_z\,} \sim
  A^3 \alpha^3 \gamma^2 \cdot 
  \frac{\eps^3}{\Delta^3} \cdot \frac{{\tilde C}_{ij}^{\;k}}{2}
  \;,
  \label{lim:bulk-3pt-asym}
\end{align}
where the triple $(i,j,k)$ is taken from the set 
$N(x,\eps){\times}N(y,\eps){\times}N(z,\eps)$. If this set contains any
triples allowed by the fusion rules, we assume that $(i,j,k)$ is
chosen to be one of them.

Demanding the bulk three point functions to have a finite limit leads
to 
\begin{align}
  A^3 \alpha^3 \gamma^2  =
  \text{(const)} \cdot \frac{\Delta^3}{\eps^3}\;.
  \label{lim:bulk-3pt-cond}
\end{align}

\subsubsection{Collection of results}

In order for the unit disc one point function and the bulk two and
three point functions to have a finite limit as $\Delta\rightarrow 0$,
we found the conditions 
\eqref{lim:bulk-2pt-cond}, \eqref{lim:unit-1pt-cond} and
\eqref{lim:bulk-3pt-cond}: 
\begin{align}
  A^2\,\alpha^2\,\gamma^2 = \frac{\Delta}{\eps^2} \;,\quad
  A\,\alpha\,\gamma = G \cdot \frac{\Delta^{1/2}}{\eps} \;,\quad
  A^3\,\alpha^3\,\gamma^2  = H \cdot \frac{\Delta^3}{\eps^3}\;,
\end{align}
for some constants $G,H$. These conditions have the unique solution
\begin{align}
  G=1 \quad,\quad
  A\,\alpha = H \cdot \Delta^2/\eps  \quad,\quad
  \gamma = 1 / H \cdot \Delta^{-3/2} \;.
\end{align}
Since only the combination $A\cdot\phi_i$ enters in the
calculation of the limiting correlators, it is not surprising that the
limit depends on the combination $A\cdot\alpha$. For convenience we
choose $\alpha=1$ and $A = H  \cdot \Delta^2/\eps$.
Different choices of $H$ correspond to different normalisations of
$M_\infty$. To reproduce the formulae in section \ref{sec:c=1}, 
one needs to choose $H=1$.

The $\Delta\rightarrow 0$ limit then yields the following
expressions:
\begin{align}
  &\text{unit disc partition function : }  
    &&Z^{\hat a} \;=\; 2^{3/4} \pi a \notag\\
  &\text{unit disc one point function : }  
    &&\bnpt{\,\phi_x(\ze,\zeb)\,}{\hat a}{disc} \;=\;  
    2^{3/4} \,(-1)^{[x]}\,\sin(\pi a x) \, (1{-}|\ze^2|)^{-x^2/2} \notag\\
  &\text{bulk two point function : }
    &&\npt{\,\phi_x(\ze,\zeb)\,\phi_y(0,0)\,} \;=\; \delta(x{-}y) 
    \cdot |\ze|^{-x^2} \;. \\
  &\text{bulk three point function : }
    &&\npt{\,\phi_x(\ze_1,\zeb_1)\,
    \phi_y(\ze_2,\zeb_2)\,\phi_z(\ze_3,\zeb_3)\,} \;=\; 
    \tfrac{1}{2}\,{\tilde C}_{ij}^{\;k} \times \notag \\
    &&&\qquad \times 
    |\ze_{12}|^{(z^2-x^2-y^2)/2}\,
    |\ze_{13}|^{(y^2-x^2-z^2)/2}\,
    |\ze_{23}|^{(x^2-y^2-z^2)/2} \notag
\end{align}
where ${\tilde C}_{ij}^{\;k}$ is understood as in
\eqref{lim:bulk-3pt-asym}. 

As a consistency check we can verify that the decomposition of
boundary states \eqref{lim:bc-sup} holds for the one point
functions. If we repeat the calculation 
in \eqref{lim:one-point-average} for the general
boundary condition $(b,c)$ instead of $(a,1)$ we find
\begin{align}
  \bnpt{\,\phi_x(\ze,\zeb)\,}{(b,c)}{disc} = 
  2^{3/4} \, (-1)^{[x]} \; \frac{\sin(\pi b x) \sin(\pi c x)}{
  \sin(\pi x)} \; (1{-}|\ze^2|)^{-x^2/2}
  = \sum_{a\in b\otimes c} \bnpt{\,\phi_x(\ze,\zeb)\,}{\hat a}{disc}\;,
\end{align}
in accordance with \eqref{lim:bc-sup}.

In the following two sections the factor of
a half in \eqref{lim:bulk-3pt-asym} is justified and it is shown
how analytic expressions can be obtained for the $\Delta\rightarrow 0$
limit of ${\tilde C}_{ij}^{\;k}$.


\subsection{Fusion rules}
\label{sec:fusion}

In the following we want to
estimate how many triples of Kac-labels in
the product $N(x,\eps)\times N(y,\eps)\times N(z,\eps)$ are allowed by
fusion. It is argued that, depending on $x,y,z$ and for $\eps$
sufficiently small, either no triples are allowed, or exactly half of
them.

The minimal model fusion rules state that $(r,s)\times(u,v)$ can fuse
to $(a,b)$ if and only if
\begin{align}
  \mathcal{N}_{ru}^{a}(p)\cdot\mathcal{N}_{sv}^{b}(p{+}1)
  + \mathcal{N}_{ru}^{p{-}a}(p)\cdot\mathcal{N}_{sv}^{p{+}1{-}b}(p{+}1)
  = 1 \;, \label{lim:fuse} 
\end{align}
where
\begin{align}
  \mathcal{N}_{ab}^c(r) = \begin{cases} 1 :& 
  |a{-}b|<c<\min(a{+}b,\,2r{-}a{-}b) \;,\; 
  a{+}b{+}c \text{ odd} \notag\\ 0 :& \text{otherwise} \end{cases}
\end{align}
In order to see what is happening in the $\eps,\Delta\rightarrow0$
limit it is helpful to reformulate these rules somewhat.

Recall the parametrisation of the set $N(x,\eps)$ as given in
\eqref{lim:Nnew}. There it was found that, for a given $x=[x]{+}f_x$, 
\begin{align}
  x_{rs} \in [x,x{+}\eps] \quad\Leftrightarrow\quad 
  r = [x]{+}s \text{ and }
  s\cdot\Delta = f_x{+}\Delta\tfrac{[x]}{2}{+}\nu_x
  \text{ where } \nu_x \in [0,\eps] \;.
\end{align}
Using these relations, we can eliminate $r,u,a$ from equation
\eqref{lim:fuse}, yielding expressions in $x,y,z$ and $s,v,b$. Then
we multiply the inequalities by $\Delta$ and neglect terms of
order $\Delta$ and $\eps$. Consider the first term in
\eqref{lim:fuse}. Applying this procedure,
$\mathcal{N}_{ru}^{a}(p)=1$ is equivalent
to saying, up to terms of order $\Delta$ and $\eps$,
\begin{align}
  |f_x{-}f_y| < f_z < \min(f_x{+}f_y\,,\,2{-}f_x{-}f_y)
  \text{ and } [x]+[y]+[z]+s+v+b \text{ odd } \;.
\end{align}
Similarly, $\mathcal{N}_{sv}^{b}(p{+}1)$ is nonzero if and only if
\begin{align}
  |f_x{-}f_y| < f_z < \min(f_x{+}f_y\,,\,2{-}f_x{-}f_y)
  \text{ and } s+v+b \text{ odd }\;. \label{lim:fuse-even}
\end{align}
It follows that the first summand in \eqref{lim:fuse} can be nonzero
only if $[x]{+}[y]{+}[z]$ is even. If this is the case, the indices 
$s,v,b$ have to be adjusted, s.t.\ $s{+}v{+}b$ is odd. If 
$|N(x,\eps)|\sim\eps/\Delta$ is
sufficiently large, i.e.\ $\Delta$ sufficiently small, then this
condition will in good approximation be true for half the triples in 
$N(x,\eps) \times N(y,\eps) \times N(z,\eps)$.

For the second summand in \eqref{lim:fuse} we find in the same way,
that it is nonzero only if
\begin{align}
  |f_x{-}f_y| < 1{-}f_z < \min(f_x{+}f_y\,,\,2{-}f_x{-}f_y)
  \text{ and } &[x]+[y]+[z] \text{ odd,} \notag\\ &
  s+v+b+p \text{ even} \;.
  \label{lim:fuse-odd}
\end{align}
The two conditions \eqref{lim:fuse-odd} and \eqref{lim:fuse-even}
give rise to the step function $P(x,y,z)$ given in \eqref{c=1:P}.

The statement for the fusion rules is
that, if $x,y\notin\Zbb$, then we can choose $\eps$ and $\Delta$
small enough, s.t.\ $P(x,y,z)=0$ implies that fields of 
$N(x,\eps) \times N(y,\eps)$ cannot fuse 
to fields in $N(z,\eps)$. Conversely, if
$P(x,y,z)=\frac{1}{2}$, in the limit $\Delta\rightarrow 0$, 
$\eps\rightarrow 0$
while letting $\eps/\Delta$ go to infinity, the number of allowed
triples will tend to a half.


\subsection{Bulk structure constants}
\label{sec:mm-bulk-sc}

In this section we take the $\Delta\rightarrow 0$ limit of the
structure constants ${\tilde C}_{ij}^{\;k}$, where we assume that
${N_{ij}}^k=1$, i.e.\ the triple is allowed by the fusion rules.

Define, as before 
in the minimal model $M_p$ the
abbreviations $d_{rs}=r{-}st$, where\footnote{
  The formula \eqref{lim:DFsc} for the bulk structure constants also
  holds for non-unitary minimal models $M(p,q)$. In this case one
  simply takes $t=p/q$.}
$t=p/(p{+}1)$.
The minimal model structure constants found by Dotsenko and Fateev in
\cite{DFa84,DFa85a,DFa85b} are
\begin{align}
  C_{(rs)(uv)}^{\hspace{20pt}(ab)} &= 
     t^{ 4(k{-}1)(\ell{-}1) } \cdot
     \prod_{m=1}^{k-1} \prod_{n=1}^{\ell-1}  
       \Big\{ d_{mn} (d_{mn}{-}d_{rs})(d_{mn}{-}d_{uv}) 
       (d_{mn}{+}d_{k\ell}{-}d_{11}{-}d_{uv}{-}d_{rs}) \Big\}^{-2}
  \notag\\
  & \quad \times \; \prod_{m=1}^{k-1} \frac{
     \G{m/t} \G{(m{-}d_{rs})/t} \G{(m{-}d_{uv})/t} 
       \G{(d_{11}{+}d_{rs}{+}d_{uv}-d_{k\ell}-d_{m\ell})/t}}{
     \G{1{-}m/t} \G{(d_{rs}{-}d_{m1})/t} \G{(d_{uv}{-}d_{m1})/t} 
       \G{(d_{k\ell}{+}d_{m\ell}{-}d_{rs}{-}d_{uv}{-}d_{11}{+}t)/t}}
  \notag\\
  & \quad \times \; \prod_{n=1}^{\ell-1} \frac{
     \G{nt} \G{d_{rs}{+}nt} \G{d_{uv}{+}nt}
       \G{d_{k\ell}{+}d_{kn}{-}d_{rs}{-}d_{uv}{-}d_{11}}}{
     \G{1{-}nt} \G{d_{1n}{-}d_{rs}} \G{d_{1n}{-}d_{uv}}
       \G{d_{rs}{+}d_{uv}{+}d_{11}{-}d_{k\ell}{-}d_{kn}{+}1}}
  \label{lim:DFsc}
\end{align}
where $k=(r{+}u{-}a{+}1)/2$, $\l=(s{+}v{-}b{+}1)/2$.
The indices have to satisfy the following fusion constraints:
\begin{align}
  &|r{-}u|+1 \le a \le \min(r{+}u{-}1,2p{-}r{-}u{-}1) 
  &&\text{ and } r{+}u{-}a{+}1 \in 2\mathbb{Z}
  \notag\\
  &|s{-}v|+1 \le b \le \min(s{+}v{-}1,2q{-}s{-}v{-}1) 
  &&\text{ and } s{+}v{-}b{+}1 \in 2\mathbb{Z}
  \;.
\end{align}
As given in \eqref{lim:DFsc}, the structure constants do not obey the
normalisation condition \eqref{lim:Mp-fixed}, 
i.e.\ in general $C_{(rs)(rs)}^{\hspace{20pt}(11)}\neq 1$. To change this one
could rescale the fields in such a way that
\begin{align}
  C_{(rs)(uv)}^{\hspace{20pt}(ab)} \rightarrow 
  \tfrac{\sqrt{C_{(ab)(ab)}^{\hspace{20pt}(11)}}}{
  \sqrt{C_{(rs)(rs)}^{\hspace{20pt}(11)}\;C_{(uv)(uv)}^{\hspace{20pt}(11)}}}
  \cdot C_{(rs)(uv)}^{\hspace{20pt}(ab)} \;.
  \label{lim:renorm}
\end{align}
This is done explicitly in \cite{DFa85b}. In the present case
this will however not be necessary, since for
\clim, the prefactor turns out to approach one:
$\lim_{t\rightarrow 1}C_{(rs)(rs)}^{\hspace{20pt}(11)}= 1$. In the
following we will thus continue to work with the original expression 
\eqref{lim:DFsc}.

The limiting procedure relies on two observations by Dotsenko
\cite{Dp.c.}. 

Firstly, the numbers $C_{(rs)(uv)}^{\hspace{20pt}(ab)}$ 
are continuous in
the parameters $d_{rs}$, $d_{uv}$ and $d_{ab}$, even though
this is not manifest in
the form \eqref{lim:DFsc}, because the parameters $r,s$, etc.\ still
enter in the ranges of the products.

Secondly, in principle one would have to carry out the 
$\Delta\rightarrow 0$
limit with the full expression \eqref{lim:DFsc}.
But the calculation can be significantly
simplified if we use Dotsenko's observation \cite{Dp.c.} 
that the full structure constants
can be obtained by analytic continuation from the $(r,1)$--subalgebra
\begin{align}
  C_{(r1)(u1)}^{\hspace{20pt}(a1)} =
  \prod_{m=1}^{k-1} \frac{
     \G{\tfrac mt} \G{\tfrac 1t (m{-}d_{r1})} \G{\tfrac 1t(m{-}d_{u1})} 
       \G{\tfrac 1{2t}(d_{r1}{+}d_{u1}{+}d_{a1}{+}1{+}t{-}2m)}}{
     \G{1{-}\tfrac mt} \G{\tfrac 1t(d_{r1}{-}d_{m1})} \G{\tfrac 1t(d_{u1}{-}d_{m1})} 
       \G{\tfrac 1{2t}({-}d_{r1}{-}d_{u1}{-}d_{a1}{-}1{+}t{+}2m)}}
  \label{lim:r1-sub}
\end{align}
The idea is to treat $d_{r1},d_{u1},d_{a1}$ as the fundamental
variables and analytically continue in those. Again, this cannot be
done as long as they appear (through $k$) in the range of the
product. To remove this obstacle, rewrite $\Gamma(x)=\exp(\ln\Gamma(x))$ and
use an integral formulation for $\ln \Gamma(x)$. The product in 
\eqref{lim:r1-sub} then becomes a sum which can be carried out
explicitly. 

In more detail, we use that, for $x>0$, \cite{GRy}(8.341.6)
\begin{align}
  \ln \G{x} = I[\; \alpha \beta^x + x\beta + \gamma \;]
\end{align}
where
\begin{align}
  I[\,f\,] = \int_0^1 \frac{\D\beta}{\beta(-\ln \beta)} f(\beta) \quad,
  \qquad \alpha = \frac{1}{1-\beta} \quad, \qquad \gamma=-\beta(\alpha{+}1) \;.
\end{align}
To carry out the sum one can use
$\sum_{m=1}^k q^m = (q^k-q)/(q-1)$. The result is 
\begin{align}
  \ln C_{(r1)(u1)}^{\hspace{20pt}(a1)} =&
  I\Big[ \frac{\alpha}{1{-}\beta^{1/t}} \Big\{ 
  \beta^{1/t}\big(1{+}\beta^{-d_{r1}/t}{+}\beta^{-d_{u1}/t}{+}\beta^{d_{a1}/t}\big)
  +\beta\big(1{+}\beta^{d_{r1}/t}{+}\beta^{d_{u1}/t}{+}\beta^{-d_{a1}/t}\big)
  \notag\\
  & \quad - \beta^{\tfrac{1+t}{2t}} \hspace{-10pt} \sum_{\eps_r,\eps_u,\eps_a=\pm 1} 
  \beta^{\tfrac{1}{2t}(\eps_r d_{r1}+\eps_u d_{u1}+\eps_a d_{a1})}
  \Big\} 
  + \beta \cdot \tfrac{1{-}t}{t} \big(d_{r1}{+}d_{u1}{-}d_{a1}{-}1{+}t\big) \Big]
  \label{lim:lnC}
\end{align}
where the integral converges for
\begin{align}
  -t<d_{r1},d_{u1}<1 \quad,\quad -1<d_{a1}<t \quad,\quad
  |d_{r1}|+|d_{u1}|+|d_{a1}| < 1+t \;.
  \label{lim:lnCrange}
\end{align}
In the formulation \eqref{lim:lnC} it is trivially possible to carry
out the analytic continuation in $d_{r1},d_{u1},d_{a1}$
by just substituting any set of real
values in the ranges \eqref{lim:lnCrange}. 
Within this range, one can now verify (at least
numerically) Dotsenko's result that the full structure constants
\eqref{lim:DFsc} can be obtained by continuation from the subalgebra
\eqref{lim:r1-sub}. 
The more precise statement is that \eqref{lim:DFsc} and 
\eqref{lim:lnC} agree in the given range once they are both normalised 
to ${C_{kk}}^1=1$ through the use of \eqref{lim:renorm}.

We also immediately see how \eqref{lim:lnC} turns into \eqref{c=1:Q}
when substituting $t=1$.


\section{Conclusion}

We have considered the limit of the diagonal minimal models $M_p$ as
$p\rightarrow\infty$. The main aim was to find a bulk theory for the
$c=1$ boundary conditions considered in an earlier work \cite{GRW01},
which were equally obtained in the $p\rightarrow\infty$ limit of minimal
models. The expression for the cylinder partition functions with these
boundary conditions implied that the bulk spectrum had to be a
continuum. Thus the limit of the minimal model bulk theory was constructed
in such a way that it yielded a continuous spectrum of bulk fields. 
Intriguingly, due to ``hidden''
analytic properties of the minimal model bulk structure constants the
limit can be given explicitly. 

The limit of expressions of minimal models led us to conjecture a
non-rational CFT with $c=1$ whose conformal boundary conditions
include the set found in \cite{GRW01}. The main argument presented to
support this conjecture was crossing symmetry of the two point
function on the upper half plane and the four point function on the
full complex plane. In the general case these were tested numerically,
employing a recursive method due to Zamolodchikov to evaluate
the conformal blocks at $c=1$. In the case where all fields in the
bulk four point function have half-integer labels the structure
constants simplify considerably and additionally analytic expressions
for the conformal four point blocks can be obtained. In the easiest
case we presented an explicitly crossing symmetric expression for the
bulk four point function.

There are however many remaining open questions. The first is the
precise relation of the present theory to Liouville theory, to which
it bears many superficial resemblances. Related to this point one may
wonder if there is an action that describes the present theory and if
in any way it can be understood as a quantisation of Liouville theory
at $c=1$. This will also help to find a physical interpretation of the
$M_\infty$ theory.

Rather than having to rely on numerical checks it would be interesting
to see if the crossing of the $c=1$ blocks can be described by fusion
matrices, as was done in the case of Liouville theory in
\cite{PTe99,Tes01}. The properties of the fusion matrices could then
be used to prove crossing symmetry.

A conformal field theory has to fulfill more consistency conditions
than just the two instances of crossing symmetry tested here
\cite{Son88,Lew92}. Other constraints come from the four point
function on the boundary, correlators involving two boundary and one
bulk field, as well as one point functions on torus and cylinder. The
first of these additional constraints was already conjectured to hold
in \cite{GRW01}. To test the second, the coupling of the $c=1$ bulk
fields to arbitrary boundary fields (and not just the identity) has to
be worked out.  This would also allow one to verify the consistency of
the limit of the bulk--boundary coupling constants with the
decomposition of the boundary conditions at $c=1$.  We leave these
problems for future research.

One may also wonder what the interpretation of $M_\infty$
would be in the context of string theory, where one could replace a
free boson describing a particular dimension in target space by
$M_\infty$. From this point of view it would be particularly
interesting to repeat the calculation presented here in the case of
$N=1,2$ super minimal models.

{\bf Acknowledgments} -- The authors would like to thank Vl.~Dotsenko
for explaining his work on analytic continuation of the minimal model
structure constants. We are further grateful to 
M.~Gaberdiel,
K.~Graham,
B.~Ponsot,
A.~Recknagel,
D.~Roggenkamp, 
V.~Schomerus,
C.~Schweigert and
J.~Teschner for helpful discussions.

GMTW would like to thank LPTHE for hospitality while part of this
research was carried out.

This work is partly supported by EU contracts HPRN-CT-2000-00122 and
HPMF-CT-2000-00747.


\setcounter{section}{1}
\setcounter{subsection}{0}
\renewcommand{\thesection}{\Alph{section}}
\section*{Appendix}

\subsection{Relation to Liouville structure constants}
\label{sec:liouv}

Here we compare in some detail the formula for the \Moo\ structure
constants \eqref{c=1:sc} to the corresponding expression in Liouville
theory, given by Dorn and Otto in \cite{DOt94} and by 
Zamolodchikov and Zamolodchikov in \cite{ZZa95}. The
following expressions are taken from \cite{ZZa95}.

Let $Q=b+1/b$. The Liouville theory has central charge 
$c_L=1+6Q^2$ and the exponential Liouville operators 
$V_\alpha(\zeta)=e^{2\alpha \phi(\zeta)}$ have conformal weight 
$\Delta_\alpha = \alpha (Q-\alpha)$. Define functions $\Upsilon(x)$ as
\begin{align}
  \log\Upsilon(x)=
  \int_0^\infty \frac{\D t}{t}\left[\left(\frac{Q}{2}-x\right)^2e^{-t}
  -\frac{\sinh^2\left( \frac{Q}{2}-x\right)\frac{t}{2}}
        {\displaystyle\sinh\frac{bt}{2}\sinh\frac{t}{2b}}
  \right]
  \label{liouv:Ups}
\end{align}
The Liouville structure constants are then:
\begin{align}
  &C(\alpha_1,\alpha_2,\alpha_3) = \left[\pi\mu\gamma(b^2)b^{2-2b^2}
  \right]^{(Q-\alpha_1-\alpha_2-\alpha_3)/b}\cdot
  \frac{\D\Upsilon(x)}{\D x}\Big|_{x=0} \times \notag\\
  &\qquad
  \frac{\Upsilon(2\alpha_1)\Upsilon(2\alpha_2)\Upsilon(2\alpha_3)}{
  \Upsilon(\alpha_1{+}\alpha_2{+}\alpha_3{-}Q)
  \Upsilon(\alpha_1{+}\alpha_2{-}\alpha_3)
  \Upsilon(\alpha_2{+}\alpha_3{-}\alpha_1)
  \Upsilon(\alpha_3{+}\alpha_1{-}\alpha_2)}
  \label{liouv:sc}
\end{align}
From the expression for the central charge $c_L=1+6Q^2$ we see that
for $c=1$ we need $Q=0$. To achieve this we set $b=i$. The
conformal weights are then given by $\Delta_\alpha = -\alpha^2$. To
bring them to the form $h_y=y^2/4$ used here let $\alpha = i y / 2$.

As a next step some factors are dropped from
\eqref{liouv:sc} which can be absorbed by re-defining the fields.
That is, we consider only the expression
\begin{align}
  \tilde C =
  \frac{\Upsilon(2\alpha_1)\Upsilon(2\alpha_2)\Upsilon(2\alpha_3)}{
  \Upsilon(\alpha_1{+}\alpha_2{+}\alpha_3)
  \Upsilon(\alpha_1{+}\alpha_2{-}\alpha_3)
  \Upsilon(\alpha_2{+}\alpha_3{-}\alpha_1)
  \Upsilon(\alpha_3{+}\alpha_1{-}\alpha_2)} 
  = \exp\big(\int_0^\infty \frac{\D t}{t} F(t) \big)\;.
  \label{liouv:reduced}
\end{align}
To obtain the function $F(t)$ we substitute all expressions 
\eqref{liouv:Ups} and rewrite $\tilde C$ as the
exponential of a single integral. In doing so we first notice that all terms 
$x^2 e^{-t}$ in the integrand of \eqref{liouv:Ups} cancel. For each
term $\Upsilon(iy)$ in \eqref{liouv:reduced} we thus get a contribution
\begin{align}
  -\big(\sinh(-iy t/2)\big)^2 \big( \sinh(it/2) \sinh(-it/2) \big)^{-1}
  = \frac{e^{it}}{(1-e^{it})} ( e^{it\,y} + e^{it(-y)} -2 )   
\end{align}
to $F(t)$. If we now modify also the integration contour in
\eqref{liouv:reduced} to run along the positive imaginary axis, 
i.e.\ we replace $t=i \tau$ where $\tau$ runs from $0$ to $\infty$,
the resulting expression is precisely the
one given for $Q(x,y,z)$ in \eqref{c=1:Q}.

The correspondence between \Moo\ and Liouville does however not
seem to be completely straightforward. First 
note that to match $Q(x,y,z)$ to
\eqref{liouv:reduced} we did indeed {\em replace} and not {\em rotate}
the complex integration contour, that is we did not worry about
poles. Second it is unclear how the step function $P(x,y,z)$ in
\eqref{c=1:sc} would appear from the Liouville point of view.

\subsection{Analytic continuation of the c=1 structure constants}
\label{sec:cont}

Recall the definition of $Q(x,y,z)$ in the bulk structure constants
\eqref{c=1:sc}
\begin{align}
  Q(x,y,z) = I\big[
  \frac{\beta}{(1{-}\beta)^2} \Big\{ \;
  2+\sum_{\eps=\pm 1}\big(\beta^{\eps x}{+}\beta^{\eps y}{+}
  \beta^{\eps z} \big) -
  \hspace{-10pt}\sum_{\eps_x,\eps_y,\eps_z=\pm 1} \hspace{-10pt}
  \beta^{(\eps_x x+\eps_y y+\eps_z z)/2}
  \Big\}\; \big] \;.
  \label{cont:sc}
\end{align}
where the functional $I$ is defined as
\begin{align}
  I[f(\beta)] = \int_0^1 \frac{\D\beta}{\beta(-\ln \beta)} f(\beta)
  = \int_0^\infty \frac{\D\tau}{\tau} f(e^{-\tau}) \qquad\;; \;
  \beta = e^{-\tau} \;.
\end{align}
Here we will use the second formulation, and split the integral in two
parts
\begin{align}
  I[f] = I_1[f] + I_2[f] \qquad\text{ where }
  I_1[f] = \int_0^1 \frac{\D\tau}{\tau} f(e^{-\tau}) \;,\;
  I_2[f] = \int_1^\infty \frac{\D\tau}{\tau} f(e^{-\tau}) \;.
\end{align}
After splitting the integral in \eqref{cont:sc} in this way, one can
verify that the part $I_1$ will converge irrespective of the values of
$x,y,z$, while for $I_2$ to converge we need $|x|,|y|,|z|<1$
and $|x+y+z|<2$. Let
\begin{align}
  G(x) = I_2\big[ \frac{\beta}{(1{-}\beta)^2} \cdot \beta^x \big] \;. 
  \label{cont:G}
\end{align}
For $x>-1$ the integral converges and we can use \eqref{cont:G} as
definition for $G(x)$. For $x\le -1$, $G(x)$ can be defined by analytic
continuation in $x$.

Before doing so, let us recall some properties of the exponential
integral $E_1(x)$ and $Ei(x)$ \cite{GRy}(8.211, 8.214) 
\begin{align}
  &E_1(x) = \int_x^\infty \frac{e^{-t}}{t} \D t 
    = -\gamma-\ln x-\sum_{n=1}^\infty \frac{(-x)^n}{n\cdot n!}
  \qquad &; \; x>0 \\
  &Ei(x) = - \lim_{\eps\rightarrow 0+} \Big(
    \int_{-x}^{-\eps} \frac{e^{-t}}{t} \D t +
    \int_{\eps}^{\infty} \frac{e^{-t}}{t} \D t \Big)
    = \gamma+\ln x+\sum_{n=1}^\infty \frac{x^n}{n\cdot n!}
  \qquad &; \; x>0 \label{cont:Ei}\\
  &E_1(-x+i\cdot 0) = -Ei(x)-i\pi \label{cont:E1-Ei}
\end{align}
Equation \eqref{cont:E1-Ei}
describes the behaviour of $E_1(x)$ under analytic continuation along
a path that passes above the origin.

Let us, for definiteness, fix the path along which we want to continue
$G(x)$ to also pass just above the negative real axis and the origin.

Note that, for $x>0$, $I_2[\beta^x]=E_1(x)$. One can quickly verify
that the following functional relation holds for $G(x)$, $x>-1$:
\begin{align}
  G(x) = 2G(x{+}1) - G(x{+}2) + E_1(x{+}1)
  \label{cont:fun-rel}
\end{align}
Upon analytic continuation the relation \eqref{cont:fun-rel} will
remain true, so that for $x\le -1$ we can use it to bring the argument
of $G$ into the positive domain. The analytic continuation of $E_1$ is
just \eqref{cont:E1-Ei}. 

One can now solve this relation explicitly. Write $x<0$ as $x=-n+f$
where $n\in\Zbb_{\ge 0}$ and $0\le f <1$. Then 
\begin{align}
  G(x) = (n{+}1)\,G(f) - n\,G(f{+}1) + n\,E_1(f) - 
  \sum_{k=1}^{n-1} (n{-}k)\,Ei(k{-}f) - \frac{n(n{-}1)}{2} i \pi
  \label{cont:G-expl}
\end{align}
The function $G(x)$ thus has poles for negative integer values of
$x$. 

With the explicit form \eqref{cont:G-expl}, the expression for
$Q(x,y,z)$ which is valid for all values of $x,y,z\in\Sbb$ is
\begin{align}
  Q(x,y,z) &= I_1\big[
  \frac{\beta}{(1{-}\beta)^2} \Big\{ \;
  2+\sum_{\eps=\pm 1}\big(\beta^{\eps x}{+}\beta^{\eps y}{+}
  \beta^{\eps z} \big) -
  \hspace{-10pt}\sum_{\eps_x,\eps_y,\eps_z=\pm 1} \hspace{-10pt}
  \beta^{(\eps_x x+\eps_y y+\eps_z z)/2}
  \Big\}\; \big] \notag \\
  &\quad + 2 G(0) + \sum_{\eps=\pm1} \big( 
  G(\eps x) + G(\eps y) + G(\eps z) \big) -  
  \hspace{-10pt}\sum_{\eps_x,\eps_y,\eps_z=\pm 1} \hspace{-10pt}
  G((\eps_x x+\eps_y y+\eps_z z)/2)
  \;.
  \label{cont:sc-cont}
\end{align}
Recall that above we made an
explicit choice for the path along which $G(x)$ was
analytically continued. $G(x)$ is singular for negative integer values
of $x$, and the path passed above these
poles. Eqn.~\eqref{cont:G-expl} shows that the singularities are all
logarithmic, so that a change of path can lead to an addition or
subtraction of an integer multiple of $2\pi i$ in
\eqref{cont:G-expl}. Since in the end it
is $e^Q$ that matters, we see that the structure constants are
independent of the path chosen for analytic continuation.

The imaginary part of $G(x)$, and thus that of $Q(x,y,z)$, is always
of the from $i m$, where $m\in\Zbb$. This implies that $c(x,y,z)$
as defined in \eqref{c=1:sc} is always real. It could take any value
in $\Rbb$, though. However one can verify (at least numerically)
that, with the choice of
contour for $G(x)$ as in \eqref{cont:G-expl},
\begin{align}
  \text{Im}\; Q(x,y,z) \neq 0 \quad \Rightarrow \quad P(x,y,z) = 0
  \;.
\end{align}
In particular this implies that $c(x,y,z)\ge 0$.


\subsection{Exact values of the $c=1$ structure constants}
\label{sec:exactcs}

In the special case that $x$ and $y$ (say) are both half integer, 
the structure constants take the simple form
\begin{align}
  c(m+\hf,n+\hf,z)
= 2^{-z^2-1}\, q_{m,n}(z)
\;,
\end{align}
where $q_{m,n}(x)$ is a polynomial 
of degree $(m+m^3+n+n^3)$.
We conjecture that this polynomial is
\begin{align}
  q_{m,n}(x) &= 
\left[
 \prod_{k=-m-n+1}^{m+n-1}
 \frac  {    (k - x)^{m + n  + 1 - |k|} }
        {    ( |k| + \epsilon )^{m + n  +1 - |k|-\epsilon} }
\right]
\!\times\!
\left[
 \prod_{k=2-|m-n|}^{|m-n|-2}
  \left( \frac{k-x}{m+n+1-k} \right)^{|m-n| - |k|}
\right]
\;,
\label{exact:q}
\end{align}
where $\epsilon=0$ is $m+n$ is even and $1$ if $m+n$ is odd, and in
both products $k$ runs over every other integer.  We have checked this
for many cases, but cannot prove that this is indeed the correct
result.  The various factors can be found by differentiating the
expression \eqref{cont:sc} for $Q$ with respect to $z$, in which case
the integration with respect to $\beta$ can be done to give an
expression involving the digamma function $\psi$. For $x$ and $y$
half-integral this simplifies to a rational function of $z$, which can
then be integrated to give the final form.  The overall constant term
however we have had to fix numerically.


\subsection{Virasoro representations at $c=1$}
\label{sec:repn}

The commutation relations for the Virasoro algebra at central charge
$c=1$ are
\begin{align}
  [L_m, L_n] = (m{-}n) L_{m+n} + \delta_{m+n,0}\tfrac{1}{12}(m{-}1)m(m{+}1) \;.
\end{align}
A highest weight state $\is{h}$ is defined by the property
\begin{align}
  L_0 \is{h} = h \is{h}
\;,\;\;\;\;
  L_m \is{h} = 0\;,\;\; m>0
\;.
\end{align}
The Verma module $M(h)$ of weight $h$ is generated by the action of
the Virasoro algebra on a highest weight state $\is{h}$ and 
has a basis of states
\begin{align}
  L_{-n_1} L_{-n_2} \dots L_{-n_k} \is{h} \qquad \text{ where }
  n_1\ge n_2\ge \dots \ge n_k \ge 1 
\;.
\label{repn:L-basis}
\end{align}
Hence, the character of $M(h)$ is 
\begin{align}
 \chi_{M(h)}(q)
= \text{tr}_{M(h)}(\, q^{L_0 - c/24} \,)
= q^h / \eta(q)
\;,
\end{align}
where $\eta(q)=q^{1/24} \prod_{n\in\Zbb_{\ge 0}} (1-q^n)$ is the
Dedekind $\eta$--function.

The irreducible representation $L(h)$ of weight $h$ is the quotient of
$M(h)$ by its maximal proper submodule.
The structure of Virasoro Verma modules is known \cite{FFu83}, 
and in particular, for $c=1$ a Verma module is irreducible unless
$h=n^2/4$ for an integer $n$.
If $h=n^2/4$ then the maximal submodule is generated by a single
highest weight vector at level $|n|+1$ and is isomorphic to the Verma
module $M(|n|+2)$, so that it is straightforward to write down the
characters $\chi_x(q)$ of irreducible highest weight representations
$L(h_x)$ with weight $h_x=x^2/4$:
\begin{align}
  \chi_x(q) = 
  \begin{cases}
    \vartheta_x(q)                          &; x \notin \Zbb \\
    \vartheta_x(q) - \vartheta_{|x|+2}(q) \;&; x \in \Zbb 
  \end{cases}
  \quad\text{ where }
  \vartheta_x(q) = \frac{q^{x^2/4}}{\eta(q)} \;.
\end{align}
We will also need the modular transformation
properties of $\vartheta_x(q)$. Let
$\text{Im}\,\tau>0$, $q=e^{2\pi i\tau}$ and $\qt = e^{-2\pi i/\tau}$.
The modular transformation of the $\eta$--function is simply
$\eta(\qt)=\sqrt{-i\tau}\,\eta(q)$. Using an inverse
Laplace transformation\footnote{
  More specifically \cite{GRy}(17.13.114):
  $(\pi/s)^{1/2} e^{-a/s} = 
  \int_0^\infty x^{-1/2} \cos(2\sqrt{ax}) e^{-sx} \D x$ where 
  $\text{Re}(s)>0$.}
one can then check that
\begin{align}
  \vartheta_y(q) = \tfrac{1}{\sqrt{2}} \cdot 
  \int_0^\infty (e^{\pi i xy}+e^{-\pi i xy}) 
  \cdot \vartheta_x(\qt) \; \D x\;.
  \label{repn:mod}
\end{align}
In the main text this relation is used in the form
\begin{align}
  \int_0^\infty \hspace{-10pt}\D x\; 2^{3/2} \cdot
  \sin(\pi a x) \cdot \sin(\pi b x)
  \cdot \vartheta_x(\qt) 
  \;=\; \vartheta_{a-b}(q) -  \vartheta_{a+b}(q) 
  \;=\; \sum_{k=|a-b|{+}1\,,\,2}^{a{+}b{-}1} \chi^{\vphantom{\phi}}_{k{-}1}(q) 
\;,
\end{align}
where in the final sum $k$ is increased in steps of two.


\subsection{Zamolodchikov's recursive methods}
\label{sec:recur}

In \cite{Zam84} and \cite{Zam87a} Al.~Zamolodchikov investigated the
analytic structure of the conformal blocks as functions of the central
charge $c$. This leads to the formulation of a recursion relation.

Consider a four point block with conformal weights $h_1, h_2, h_3,
h_4$ and internal channel $H$, with the two insertions taken at the
coordinates $1$ and $\ze$
\begin{align}
  \bL{\cbB{h_1}{h_2}{h_3}{h_4\hspace{-5pt}}H1\ze} 
= \ze^{H-h_3-h_4} F(c,h_i,H,\ze)
  \;. 
  \label{cross:generic-H-block}
\end{align}
Here $c$ is the central charge.
In this definition of the function $F$ the divergence for
$\ze\rightarrow0$ has been extracted and $F$ now has the 
expansion 
$F=1+a_1\ze+a_2\ze^2+\dots$. 
In order to work out the coefficient of
$\ze^n$ we have to sum over all intermediate states at level $n$. The
function $F$ then takes the form
\begin{align}
   F(c,h_i,H,\ze) = 1 + \sum_{n=0}^\infty
   \frac{P_n(c,h_i,H)}{Q_n(c,H)} \cdot \ze^n \;.
   \label{cross:zam-ze-series}
\end{align}
The factor $P_n$ is a polynomial in $c,h_i,H$ and originates from
evaluating the three point functions for an insertion of states of the
form 
$L_{-m_1}\dots L_{-m_k}\is{H}\os{H}L_{n_1}\dots L_{n_\l}$. 
The factor
$Q_n$ is a polynomial in $c,H$ and results from computing the inverse
of the inner product matrix of level $n$ vectors in the highest
weight representation
$L(c,H)$. Consequently it is independent of the $h_i$.

For generic values of $c$ and $H$, the highest weight representation
$L(c,H)$ is just the Verma module $M(c,H)$. 
The determinant of its inner product matrix
at level $n$ is by definition the Kac-determinant (see \cite{KRa87}
for more details). 

If we understand $F(c,h_i,H,\ze)$ as a function of $c$, 
equation \eqref{cross:zam-ze-series} shows that its possible
poles are located at zeros of $Q(c,H)$, i.e.\ at values of $c$ where
$M(c,H)$ becomes degenerate. For fixed $H$, the Kac determinant
tells us that $V(c,H)$ has null vectors when ever $c$ takes one of the
values $c_{mn}(H)$, defined as follows
\begin{align}
  c_{mn}(H) &= 13 - 6\big(t_{mn}(H)+(t_{mn}(H))^{-1}\big) \;,
\intertext{where}
  t_{mn}(H) &= \frac{2H+mn-1+\sqrt{4H(H+nm-1)+(m-n)^2}}{n^2-1} > 0
  \label{cross:zam-tdef} 
\end{align}
and the indices $m,n$ run over integer values $m\ge 1$ and $n \ge 2$.
Since $H\ge 0$, the argument of the square root is always
positive. Furthermore, we always choose the positive branch
of the square root, i.e.\ $\sqrt{x}>0$ for $x>0$.

We now know the location of the poles of $F(c,h_i,H,\ze)$ as a function
of $c$. Zamolodchikov argues that these are simple poles and
that the residues of the poles are again
proportional to conformal blocks and gives the expansion
\begin{align}
  F(c,h_i,H,\ze) = & {}_2F_1(H{+}h_1{-}h_2,H{+}h_3{-}h_4;2H;\ze)
  \notag\\
  &+ \sum_{m\ge 1, n\ge 2}^\infty 
  \frac{ A_{mn}(H) P_{mn}(h_i,H)}{c-c_{mn}(H)}
  \cdot \ze^{m\cdot n} \cdot
  F(c_{mn}(H),h_i,H{+}mn,\ze) \;. 
  \label{cross:generic-H-F}
\end{align}
This is the sought after recursion relation. ${}_2F_1$ denotes the
hypergeometric function.
The factor $P_{mn}(h_i,H)$ is designed to give zero in case that the
fusions $h_1\times h_2\rightarrow H$ and $h_3\times h_4\rightarrow H$
are allowed even at the degenerate point $c=c_{mn}(H)$. It is given by 
\begin{align}
   P_{mn}(h_i,H) = 
   \prod_{p,q} \Big\{\; (\l_2{+}\l_1{-}\l_{pq})\,
   (\l_2{-}\l_1{-}\l_{pq})\,(\l_3{+}\l_4{-}\l_{pq})\,
   (\l_3{-}\l_4{-}\l_{pq})\; \Big\}
\end{align}
Here the product runs over $p=-m{+}1 , -m{+}3, \dots,$ $m{-}3, m{-}1$ and
$q=-n{+}1 , -n{+}3, \dots,$ $n{-}3, n{-}1$. The $\l_i$ and
$\l_{pq}$ are all functions of $m,n$ and $H$ and are defined as
\begin{align}
  \l_i(m,n,H) = \sqrt{h_i + \frac{(1{-}t_{mn}(H))^2}{4 t_{mn}(H)}}
  \qquad, \qquad
  \l_{pq}(m,n,H) = \frac{ p-q\cdot t_{mn}(H)}{\sqrt{4 t_{mn}(H)}} \;.
  \label{cross:zam-ldef} 
\end{align}
Up to this point the formulae where more or less determined by
consideration of the analytic structure of the conformal blocks. The
astonishing fact is that Zamolodchikov managed to find an explicit
expression also for the normalisation factor $A_{mn}(H)$:
\begin{align}
  A_{mn}(H) = \frac{12\big(t-\tfrac{1}{t}\big)}{ 
  (m^2{-}1)\tfrac{1}{t} - (n^2{-}1)\,t }
  \cdot \prod_{a,b} \frac{1}{2\,\l_{ab}} \;.
\end{align}
Here it is again understood that $t=t_{mn}(H)$ and 
$\l_{ab}=\l_{ab}(m,n,H)$, as defined in \eqref{cross:zam-tdef} and
\eqref{cross:zam-ldef}. The product runs over the range
$a=-m{+}1,-m,\dots,$ $m{-}1,m$ and 
$b=-n{+}1,-n,$ $\dots,$ $n{-}1,n$. The two points $(a,b)=(0,0)$ and
$(a,b)=(m,n)$ have to be omitted from the product.

\subsubsection{Recursion for internal channel H=0}

In the test of crossing symmetry we will also need the conformal
blocks involving the degenerate internal channel $H=0$. 
In this case the sum over states in
the internal channel is taken over a smaller set, since the null
vectors in the Verma module $M(c,H{=}0)$ have to be left out. 

It turns out that the $H=0$ block is not obtained by taking the limit
$H\rightarrow 0$ of expression \eqref{cross:generic-H-block}. Instead 
one has to slightly modify the recursion formula \eqref{cross:generic-H-F}. 
Comparison with 
the explicit computation of the expansion in $\ze$ 
suggests that only the first level of the recursion is
affected. The relation we find is
\begin{align}
  \bL{\cbB{h_1}{h_2}{h_3}{h_4\hspace{-5pt}}01\ze} &\;=\; \ze^{-h_3-h_4} 
  F_{\text{null}}(c,h_i,\ze) \;, \\[10pt]
  F_{\text{null}}(c,h_i,\ze) &\;=\; 1 + \lim_{H\rightarrow 0}
  \underset{m\neq n}{\sum_{m\ge 1, n\ge 2}^\infty}
  \frac{ A_{mn}(H) P_{mn}(h_i,H)}{c-c_{mn}(H)}
  \cdot \ze^{m\cdot n} \cdot
  F\big(c_{mn}(H),h_i,H+mn,\ze\big)
  \label{cross:H=0-block}
\end{align}
This relation of course makes sense only for $h_1=h_2$ and $h_3=h_4$
since otherwise $H=0$ is forbidden. $A_{mn}$,
$P_{mn}$, $c_{mn}$ and $F$ are the same as in the previous
section. Note that in the first level of the recursion 
(i.e.\ in $F_{\text{null}}$) all diagonal
terms $m=n$ are left out, while in higher levels 
(i.e.\ in $F$) they are again included. If one tries to start the recursion
in \eqref{cross:H=0-block} directly with $H=0$, one finds that some of 
the $A_{mn}$ become singular. In taking the $H\rightarrow 0$ limit these
singularities cancel, and one obtains a finite answer.

\subsubsection{Recursion relation from pole structure in H}

The recursion relations \eqref{cross:generic-H-F} and
\eqref{cross:H=0-block}, which we will refer to as $c$--recursion,
have the advantage of directly producing the power series expansion in
$\ze$. To check its correctness,
this result can be compared term by term to the expansion resulting
from inserting a basis of intermediate states in the corresponding
block.

In \cite{Zam87a} Zamolodchikov describes an alternative method to
compute a conformal block, which derives from the pole structure in
$H$ rather than in $c$. This formulation of the blocks, which we
refer to as $H$--recursion, has better
convergence properties than the $c$--recursion, but it is less
straightforward to extract the expansion in $\ze$. 

To see if a program using these recursion relations works correctly,
one could start by implementing and verifying
the $c$--recursion, and then comparing it to the results from the
$H$--recursion. 

Below we reproduce the formulae for the $H$--recursion in the case
of $c\le 1$ and a non-degenerate internal channel $H$.
\begin{align}
  \bL{\cbB{h_1}{h_2}{h_3}{h_4\hspace{-5pt}}H1\ze} &\;=\; \ze^{H-h_3-h_4} 
  \,\mathcal{F}(c,h_i,H,\ze) \;, \notag\\[10pt]
  \mathcal{F}(c,h_i,H,\ze) &\;=\; \frac{(16 q)^{ H + (1-c)/24 }
   \cdot \mathcal{H}(c,h_i,H,q) }{
  x^{ (1-c)/24 + h_3 + h_4 } \,
  (1{-}x)^{ (1-c)/24 + h_2 + h_3 } \cdot
  \big(\theta_3(q)\big)^{ (1-c)/2 + 4(h_1+h_2+h_3+h_4) }}
  \notag\\[10pt]
  \mathcal{H}(c,h_i,H,q) &\;=\; 1 + \sum_{m\ge 1, n\ge 1}
     \frac{ (16 q)^{m\cdot n} R_{mn}(c,h_i)}{H-H_{mn}(c)}
     \mathcal{H}(c,h_i,H_{mn}(c)+mn,q) 
  \label{cross:H-recurs}
\end{align}
Note that contrary to \eqref{cross:generic-H-F} and
\eqref{cross:H=0-block},
in this recursion the sum starts at one for both $m$ and $n$.
The theta function is 
$\theta_3(q)=\sum_{n\in\Zbb} q^{n^2}$ and 
$q$ is a function of $\ze$, given by
\begin{align}
 q(\ze) = \exp\big(i \pi \tau(\ze)\big)\quad,\quad 
 \tau(\ze) = i \cdot \frac{K(1{-}\ze)}{K(\ze)}\quad,\quad
 K(\ze) = \frac{1}{2} \int_0^1 \frac{\D t}{\sqrt{
 t(1{-}t)(1{-}\ze t)}}\;.
\end{align}
The coefficients $R_{mn}$ are
\begin{align}
   R_{mn}(c,h_i) = -\frac{1}{2} \;
   \frac{\prod_{p,q}  
   (\lambda_2{+}\lambda_1{-}\tfrac{\lambda_{pq}}{2})\,
   (\lambda_2{-}\lambda_1{-}\tfrac{\lambda_{pq}}{2})\,
   (\lambda_3{+}\lambda_4{-}\tfrac{\lambda_{pq}}{2})\,
   (\lambda_3{-}\lambda_4{-}\tfrac{\lambda_{pq}}{2})}{
   \prod_{a,b} \lambda_{ab}}
\end{align}
The product in the numerator runs over
\begin{align}
  p &= -m{+}1,-m{+}3,\dots,m{-}3,m{-}1 \;,\notag\\ 
  q &= -n{+}1,-n{+}3,\dots,n{-}3,n{-}1 \;,
\end{align} 
while the product in the denominator runs over
\begin{align}
  a &= -m{+}1,-m{+}2,\dots,m{-}1,m  \;,\notag\\ 
  b &= -n{+}1,-n{+}2,\dots,n{-}1,n \;.
\end{align} 
The two pairs $(a,b)=(0,0)$ and
$(a,b)=(m,n)$ have to be omitted from the 
product in the denominator.
$\lambda_i$ and $\lambda_{pq}$ are functions of $c$ given by
\begin{align}
  \lambda_i(c) = \sqrt{ h_i + \tfrac{1{-}c}{24} }\quad,\quad
  \lambda_{pq}(c) = \alpha_+ \cdot p + \alpha_- \cdot q \quad,\quad
  \alpha_\pm = \tfrac{1}{\sqrt{24}} \Big( \sqrt{1{-}c} \pm \sqrt{25{-}c}
  \Big) \;.
\end{align}
Finally
\begin{align}
  H_{mn}(c) = \frac{c{-}1}{24} + \frac{\big( m \cdot \alpha_+(c) + 
  n \cdot \alpha_-(c) \big)^2}{4} \;.
\end{align}

\newcommand{\bibit}[1]{\bibitem{#1}}

\end{document}